\documentclass{jfm}

\usepackage{longtable}
\usepackage{amssymb}
\usepackage{amsmath}
\usepackage{graphicx}
\usepackage{bm}
\usepackage{dcolumn}
\usepackage{booktabs}
\usepackage{color}
\usepackage{appendix}
\usepackage[export]{adjustbox}
\usepackage[section]{placeins}
\usepackage[T1]{fontenc}
\usepackage[utf8]{inputenc}
\usepackage{array}
\usepackage{capt-of}
\usepackage{epstopdf}
\usepackage{hyperref}
\hypersetup{
    colorlinks = true,
    urlcolor   = red,
    linkcolor = blue,
    citecolor = blue,
    filecolor = black,
}

\shorttitle{A DBM for thermodynamic nonequilibrium transport}
\shortauthor{Li, He, Lai, Gan, Liu and Lin}

\begin{document}

\title{A discrete Boltzmann model with state-dependent power-law relaxation time for nonequilibrium transport in compressible flows}

\author{
Demei Li\aff{1},
Zhongyi He\aff{1},
Huilin Lai\aff{1}\corresp{hllai@fjnu.edu.cn},
Yanbiao Gan\aff{2}\corresp{gan@nciae.edu.cn},
Hailong Liu\aff{3},
Pengfei Lin\aff{4,5}
}

\affiliation{
\aff{1}School of Mathematics and Statistics, Fujian Normal University, Key Laboratory of Analytical Mathematics and Applications (Ministry of Education), Fujian Key Laboratory of Analytical Mathematics and Applications (FJKLAMA), Center for Applied Mathematics of Fujian Province (FJNU), 350117 Fuzhou, China
\aff{2}Hebei Key Laboratory of Trans-Media Aerial Underwater Vehicle, North China Institute of Aerospace Engineering, 065000 Langfang, China
\aff{3}Laoshan Laboratory, 266237 Qingdao, People's Republic of China
\aff{4}State Key Laboratory of Earth System Numerical Modeling and Application, Institute of Atmospheric Physics, Chinese Academy of Sciences, 100029 Beijing, China
\aff{5}College of Earth and Planetary Sciences, University of Chinese Academy of Sciences, 100049 Beijing, China
}

\maketitle

\begin{abstract}
Thermodynamic nonequilibrium effects play a central role in momentum and energy transport in compressible flows. In conventional BGK kinetic models, the relaxation time $\tau$ is taken as a constant, which neglects the dependence of the relaxation process on local macroscopic states. To overcome this limitation, we develop a discrete Boltzmann model with a density- and temperature-dependent power-law relaxation time, termed DTRT-DBM, in which $\tau=\tau_0(\rho/\rho_0)^a(T/T_0)^b$. This formulation extends the discrete Boltzmann framework to flows with spatially varying nonequilibrium intensity. The model is validated by the Sod shock tube and by analytical solutions for viscous stress and heat flux, demonstrating accurate recovery of both macroscopic wave structures and nonequilibrium quantities across shock waves, rarefaction waves, and contact discontinuities. On this basis, phase diagrams of viscous stress and heat flux are constructed to examine how these quantities depend on the power-law exponents $a$ and $b$. The extrema of these quantities depend exponentially on the model parameters and exhibit regime-dependent behaviour. The roles of $a$ and $b$ are not symmetric: the nonequilibrium response is more sensitive to $a$ when density gradients dominate, but more sensitive to $b$ when temperature gradients dominate. Within the parameter range and flow configurations examined here, higher-order viscous stress increases the growth rate of the total viscous-stress extremum, whereas higher-order heat flux reduces the growth rate of the total heat-flux extremum. These results show that the proposed model can capture different higher-order nonequilibrium responses in compressible flows and provides a framework for the modelling and analysis of multiscale nonequilibrium processes.
\end{abstract}

%\begin{keywords} kinetic theory,
% Authors should not enter keywords on the manuscript, as these must be chosen by the author during the online submission process and will then be added during the typesetting process (see http://journals.cambridge.org/data/\linebreak[3]relatedlink/jfm-\linebreak[3]keywords.pdf for the full list)
%\end{keywords}

\section{Introduction}

Thermodynamic nonequilibrium is a defining feature of many compressible flows in extreme and high-energy-density environments. It arises from both strong spatial gradients of macroscopic fields, such as density, temperature, velocity, and composition, and the mismatch between rapid hydrodynamic evolution and finite molecular relaxation. Under shock compression, intense shear, interfacial instability, strong heating, or chemical reaction, the local state can reorganize on very short spatial and temporal scales, leaving the particle distribution far from local Maxwellian equilibrium. Meanwhile, collisional relaxation, dissipation, and energy exchange feed back on the macroscopic response. Nonequilibrium behaviour therefore results from the continuous interplay between flow evolution and local relaxation, which determines its intensity, structure, and dynamics.

Such effects are central to many applications. In hypersonic re-entry flows, strong aerodynamic heating induces molecular dissociation and vibrational relaxation within the shock layer, and the resulting thermochemical nonequilibrium directly affects aerothermal loads and thermal-protection design \citep{he2025Energies, xiong2025POF, varma2025JFM, steer2026JFM, cao2024AST}. In combustion and detonation, nonequilibrium interactions between chemical kinetics and fluid motion govern flame propagation and detonation-wave stability \citep{li2026CAF, yu2025AET, lipkowicz2025CAF, ullman2025CAF, barwey2025CF}. Similar issues also arise in inertial confinement fusion and ablation-driven hydrodynamic instability, where nonequilibrium transport strongly influences compression performance, energy localization, and interfacial perturbation growth \citep{zhou2025ARFM, yan2025MRE, wang2009EPL, wang2010POP, wang2012POP, wang2013PS, zhang2020POP}. A predictive description of such systems therefore requires more than the recovery of macroscopic conservation laws. It also requires a model that explicitly links local relaxation to the evolving macroscopic state, so that nonequilibrium transport can be represented realistically rather than through a constant relaxation time.

Numerical simulation has become a major tool for investigating nonequilibrium fluid systems \citep{sun2019CJA, boccelli2024JCP}. By the scale of description, numerical methods may be classified as macroscopic, microscopic, or mesoscopic \citep{guo2025POF, he2025POF, wu2025POF, Xu2022BSTP, xu2025POF}. Macroscopic methods, represented by the Euler and Navier--Stokes (NS) equations, rely on the continuum assumption and near-equilibrium closure, which limits their ability to describe strongly nonequilibrium transport in shock waves, interfacial instabilities, and rarefied flows \citep{bernardini2023CPC, ding2022Energy, morgan2017JT}. Microscopic methods, such as molecular dynamics, resolve nonequilibrium processes directly through intermolecular interactions, but remain prohibitively expensive for large-scale systems \citep{mao2023PECS, klima2018JCTC, li2021JML, Xu2022BSTP}. Mesoscopic methods provide an effective compromise by offering coarse-grained kinetic descriptions that bridge the macroscopic and microscopic scales \citep{sun2025FOP}.

Representative mesoscopic approaches include the gas-kinetic scheme (GKS) \citep{xu2005JCP, sun2021ACT}, the unified gas-kinetic scheme (UGKS) \citep{zhu2017POF}, the discrete unified gas-kinetic scheme (DUGKS) \citep{guo2021AA}, the discrete velocity method (DVM) \citep{yang2019JCP, yang2022POF}, the lattice Boltzmann method (LBM) \citep{succi2018book, tran2022CF, latt2020PTRS, zhan2025JCP, wang2022AMC, fei2024JFM, yang2025US, hou2025POF}, and the discrete Boltzmann method (DBM) \citep{zhang2022POF, sun2024POF, song2024POF, song2025POF, xu2012FOP, xu2018KT, Xu2022BSTP, xu2024FOP, lai2026JFM}. Among them, LBM has been widely used for multiphase flows, reactive flows, micro- and nanoscale flows, and hydrodynamic instabilities because it is efficient, local, and flexible for complex boundaries and multicomponent systems \citep{tayyab2021POF, wei2018AMC, wang2020CMA, sawant2022JFM, huang2024POF, vienne2024POF, hosseini2024PECS}. However, LBM is designed mainly to recover macroscopic variables and their evolution. Building on the kinetic framework of LBM, DBM extends the description to thermodynamic nonequilibrium by enforcing the moment relations required for nonequilibrium modelling and retaining higher-order non-conserved moments \citep{xu2018KT, Xu2022BSTP, xu2024FOP}. Since the pioneering work of Xu \emph{et al.} \citep{xu2012FOP}, DBM has developed into a systematic framework for detecting, describing, and analysing nonequilibrium effects in complex flows. Subsequent work has extended DBM to multiphase systems, combustion, reactive flows, droplet coalescence, and hydrodynamic instabilities \citep{xu2024FOP, huang2025ATE, sun2025FOP, li2022CTP, chen2024SCP}.

DBM has also advanced towards higher-order nonequilibrium modelling. Chapman--Enskog multiscale expansion has enabled DBM models that capture second- and third-order thermodynamic nonequilibrium effects \citep{gan2018PRE}. Multiple-relaxation-time (MRT) DBM formulations with source terms have further enabled coupled descriptions of flow, reaction, and nonequilibrium effects in combustion and detonation, including three-dimensional configurations \citep{lin2019PRE, ji2021AIPA, ji2022JCP}. Even so, existing DBM formulations still do not fully recover realistic transport behaviour. In many models, the relaxation time is prescribed as a constant, so the relaxation process is decoupled from the local thermodynamic state. This simplification ignores how variations in density, temperature, and pressure affect nonequilibrium dissipation and transport, and therefore cannot capture the distinct transport responses induced by strongly non-uniform driving, such as shock heating, interfacial cooling, or ablative evaporation.

Both experiments and theory support the need for state-dependent relaxation. Millikan and White showed through vibrational-relaxation experiments that the relaxation time correlates empirically with temperature, pressure, and molecular characteristic temperature \citep{millikan1963TJCP}. More recently, Streicher \emph{et al.} showed that, in multicomponent systems, temperature evolution, pressure non-ideality, and concentration variations all significantly affect relaxation-time dynamics, especially under extreme conditions \citep{streicher2020POF}. These findings indicate that a physically meaningful kinetic model must move beyond the constant-relaxation-time assumption and establish an explicit relation between relaxation and macroscopic variables.

One way to improve physical consistency is to adopt the MRT framework, which assigns independent relaxation times to different kinetic moments. This strategy increases the flexibility of the kinetic description \citep{liang2017CMA, lin2021PRE, lin2023JPCS, chen2024SCP}. However, most MRT formulations still prescribe constant relaxation times, and MRT-DBM introduces multiple free parameters for different moment channels, whose sensitivity and selection criteria remain problem dependent and often unclear \citep{liang2017CMA, lin2021PRE, lin2023JPCS, chen2024SCP, zhang2025AAMM}. In practice, this feature often requires repeated case-by-case tuning and reduces the transparency of physical interpretation.

%A different approach has been developed within unified gas-kinetic frameworks, such as gas-kinetic unified algorithm (GKUA) and direct simulation BGK (DSBGK) method, in which the relaxation time is coupled more directly to the flow characteristics. In these methods, the relaxation time is usually related quantitatively to macroscopic variables, molecular models and local flow-regime features, allowing non-equilibrium effects to be represented more flexibly across different regimes \citep{hu2021CCP, li2020ICHMT, li2020IJNFM, wu2021JCP}. Such models can improve predictive accuracy, but their relaxation-time formulations often involve several empirical parameters and complex functional dependencies, which can obscure the separate effects of different factors on transport processes. Thus, although these models are useful for practical applications, their complexity may limit the simplicity of the model structure, the clarity of parameter interpretation and the convenience of systematic analysis.

A different approach has been developed within unified gas-kinetic frameworks, such as gas-kinetic unified algorithm  and direct simulation BGK method, in which the relaxation time is coupled more directly to flow
characteristics. In these methods, the relaxation time is usually related quantitatively
to macroscopic variables, molecular models and local flow-regime features, allowing
nonequilibrium effects to be represented across different regimes
\citep{hu2021CCP, li2020ICHMT, li2020IJNFM, wu2021JCP}. Such models have shown good
predictive capability in multiscale flows. At the same time, their relaxation-time
formulations may involve multiple parameters and functional dependencies tailored
to specific physical settings. For the present purpose, where the aim is to construct
a compact DBM formulation and to examine the separate effects of density and
temperature on nonequilibrium transport, it is useful to adopt a simpler
state-dependent form. This consideration motivates the power-law relaxation-time
model introduced below.

A related route has been developed within unified gas-kinetic frameworks, such as the gas-kinetic unified algorithm (GKUA) and the direct simulation BGK (DSBGK) method, where the relaxation time is coupled more directly to flow characteristics. In these methods, the relaxation time is usually related quantitatively to macroscopic variables, molecular models and local flow-regime features, allowing nonequilibrium effects to be represented across different regimes \citep{hu2021CCP, li2020ICHMT, li2020IJNFM, wu2021JCP}. These models have shown good predictive capability in multiscale flows. Their relaxation-time formulations may, however, involve several parameters and functional dependencies tailored to specific physical settings. For the present work, which aims to construct a compact DBM formulation and to examine the separate effects of density and temperature on nonequilibrium transport, a simpler state-dependent form is desirable. This consideration motivates the power-law relaxation-time model introduced below.

We therefore develop a discrete Boltzmann model with a density- and temperature-dependent power-law relaxation time, termed DTRT-DBM. In this model, the relaxation time is prescribed as
$\tau=\tau(\rho,T)=\tau_0\left(\rho/\rho_0\right)^a\left(T/T_0\right)^b$.
This form preserves the simplicity and analytical tractability of the BGK framework while allowing the local relaxation process to vary with the macroscopic state. It therefore provides a compact kinetic framework for describing thermodynamic nonequilibrium in compressible flows with spatially varying density and temperature.

The remainder of the paper is organized as follows. Section~\ref{II} presents the construction of the density- and temperature-dependent relaxation-time DBM. Section~\ref{III} validates the model using analytical solutions and benchmark problems. Section~\ref{IV} analyses the nonequilibrium characteristics captured by the model and examines how the state-dependent relaxation time affects different transport channels. Section~\ref{V} summarizes the main conclusions and outlines possible extensions.

%A dynamic relaxation model is therefore needed that preserves the simplicity of the BGK framework while incorporating state dependence and physical consistency in a unified form. To this end, we develop a discrete Boltzmann model with a density- and temperature-dependent power-law relaxation time, termed DTRT-DBM. In this model, the relaxation time is written as $\tau=\tau(\rho,T)=\tau_0\left(\rho/\rho_0\right)^a\left(T/T_0\right)^b$. This form allows local relaxation to respond directly to macroscopic-state variations while preserving the simplicity and analytical tractability of the BGK framework. It also provides a modelling framework for describing thermodynamic nonequilibrium in compressible flows with spatially varying local states.

%The remainder of the paper is organized as follows. Section~\ref{II} presents the construction of the density- and temperature-dependent relaxation-time DBM. Section~\ref{III} validates the model against analytical solutions and benchmark problems. Section~\ref{IV} analyses the nonequilibrium characteristics captured by the model and examines how the state-dependent relaxation time affects different transport channels. Section~\ref{V} concludes the paper.

\section{Discrete Boltzmann modelling with a state-dependent power-law relaxation time} \label{II}

To model nonequilibrium transport while retaining analytical tractability, we start from the Boltzmann equation with a simplified collision operator. A series of such operators have been proposed, including the BGK, ES-BGK, Shakhov, Rykov, and Liu models \citep{holway1966POF, shakhov1968FD, liu1990POF, rykov2008FD}. Among them, the classical BGK model approximates the collision term by assuming that the distribution function relaxes towards the local equilibrium distribution with a single relaxation time:
\begin{equation}\label{h1}
\partial_t f+\mathbf{v} \cdot \bm{\nabla} f=-\frac{1}{\tau}\left(f-f^{eq}\right).
\end{equation}
Here, $\tau$ is the relaxation time, which characterizes the rate at which the system approaches equilibrium, and $f^{eq}$ is the Maxwellian equilibrium distribution:
\begin{equation}\label{h2}
f^{eq}=\frac{\rho}{2 \pi R T}
\left(\frac{1}{2 \pi n R T}\right)^{1 / 2}
\exp \left[
-\frac{\left|\mathbf{v}-\mathbf{u}\right|^2}{2 R T}
-\frac{\eta^2}{2 n R T}
\right].
\end{equation}
Here, $\rho$, $\mathbf{v}$, $\mathbf{u}$, and $T$ denote the local density, particle velocity, flow velocity, and temperature, respectively; $R$ is the gas constant; $n$ is the number of additional degrees of freedom associated with molecular rotation or vibration; and $\eta$ is a free parameter introduced to represent these additional degrees of freedom.

In the classical BGK model, the relaxation process evolves on a fixed timescale. To describe its dependence on the local thermodynamic state, we prescribe the relaxation time in the power-law form
\begin{equation}\label{h3}
\tau=\tau(\rho, T)=\tau_0\left(\frac{\rho}{\rho_0}\right)^a\left(\frac{T}{T_0}\right)^b.
\end{equation}
Here, $\tau_0$ is the reference relaxation time, while $\rho_0$ and $T_0$ are the reference density and temperature, respectively. The exponents $a$ and $b$ determine how the relaxation time varies with density and temperature. Substituting Eq.~\eqref{h3} into the collision term yields the continuous DTRT Boltzmann equation:
\begin{equation}\label{h4}
\partial_t f+\mathbf{v} \cdot \bm{\nabla} f=-\frac{1}{\tau(\rho, T)}\left(f-f^{eq}\right).
\end{equation}
After velocity discretization, the corresponding discrete equation becomes
\begin{equation}\label{h5}
\partial_t f_i+\mathbf{v}_i \cdot \bm{\nabla} f_i=-\frac{1}{\tau(\rho, T)}\left(f_i-f_i^{eq}\right).
\end{equation}

The next step is to discretize the velocity space. Because particle velocities are continuously distributed over an unbounded phase space, whereas numerical computation can only be performed on a finite set of discrete velocities, the discrete velocity model must preserve the kinetic moments required by the target hydrodynamic description. In this sense, velocity-space discretization is the key step in DBM construction. To ensure kinetic consistency, the discrete velocity set must satisfy the moment-matching relation
\begin{equation}\label{h6}
\sum_i f_i \bm{\Psi}\left(\mathbf{v}_i, \eta_i\right)=\mathbf{M}_{m,n}=\iint f \bm{\Psi}(\mathbf{v}, \eta)\, d \mathbf{v}\, d \eta,
\end{equation}
where
\begin{equation}\label{h6.5}
\bm{\Psi}\left(\mathbf{v}_i, \eta_i\right)=\left[1, \mathbf{v}_i, \frac{1}{2}\left(v_i^2+\eta_i^2\right), \mathbf{v}_i \mathbf{v}_i, \cdots\right]^T.
\end{equation}

In practical DBM construction, the discrete equilibrium distribution is determined by requiring it to reproduce a prescribed set of equilibrium kinetic moments exactly on the finite velocity set. The corresponding equilibrium moment constraints read
\begin{equation}\label{h7}
\sum_i f_i^{eq} \boldsymbol{\Psi}^{\prime}\left(\mathbf{v}_i, \eta_i\right)=\mathbf{M}_{m,n}^{eq}=\iint f^{eq} \boldsymbol{\Psi}^{\prime}(\mathbf{v}, \eta)\, d \mathbf{v}\, d \eta.
\end{equation}
Here, $\boldsymbol{\Psi}^{\prime}\left(\mathbf{v}_i, \eta_i\right)$ denotes the set of basis functions associated with the retained higher-order moments. The number and order of these moments determine both the depth of nonequilibrium description and the order of the recovered hydrodynamic equations. For example, retaining the first five groups of kinetic moments, $\mathbf{M}_0-\mathbf{M}_{3,1}$, is sufficient to recover the Euler equations; retaining the first seven groups, $\mathbf{M}_0-\mathbf{M}_{4,2}$, allows recovery of the NS equations; and retaining the first nine groups, $\mathbf{M}_0-\mathbf{M}_{5,3}$, enables recovery of the Burnett equations. The analytical expressions of the first nine groups of kinetic moments required at the Burnett level are given in appendix~\ref{A}.

Once the required moment set is specified, the equilibrium distribution function can be obtained from the moment constraints in matrix form:
\begin{equation}\label{h8}
\mathbf{M}=\mathrm{C} \cdot \mathbf{f}^{eq},
\end{equation}
which gives
\begin{equation}\label{h9}
\mathbf{f}^{eq}=\mathrm{C}^{-1} \cdot \mathbf{M}.
\end{equation}
Here, $\mathbf{f}^{eq}=\left(f_1^{eq}, f_2^{eq}, f_3^{eq}, \cdots, f_{25}^{eq}\right)^T$ and $\mathbf{M}=\left(\mathbf{M}_0, \mathbf{M}_1, \mathbf{M}_{2,0}, \cdots, \mathbf{M}_{5,3}\right)^T=\left(M_0, M_{1x}, M_{1y}, \cdots, M_{5,3yyy}\right)^T$ is the moment vector of $f_i^{eq}$. The matrix $\mathrm{C}=\left(\mathbf{c}_1, \mathbf{c}_2, \mathbf{c}_3, \cdots, \mathbf{c}_{25}\right)$ is a $25\times 25$ coefficient matrix, where
\begin{equation}\label{h904}
\mathbf{c}_i=\left[1, v_{ix}, v_{iy}, \cdots, \frac{1}{2}\bigl(v_i^2+\eta_i^2\bigr) v_{iy} v_{iy} v_{iy}\right]^T.
\end{equation}
This inverse-matrix method guarantees the completeness of the equilibrium distribution function at the level of the retained moments and provides the basis for high-order recoverability. Since the present procedure follows the standard DBM construction, the resulting DTRT-DBM remains structurally compatible with conventional BGK-DBM while incorporating state-dependent relaxation.

With the discrete kinetic model established, Chapman--Enskog (CE) multiscale analysis can be used to examine its hydrodynamic recoverability. The introduction of $\tau(\rho,T)$ does not alter the form of the conservation laws, but it modifies the constitutive relations through the state dependence of the transport coefficients. The corresponding generalized hydrodynamic equations are
\begin{equation}\label{h10}
\partial_t \rho+\bm{\nabla} \cdot(\rho \mathbf{u})=0,
\end{equation}
\begin{equation}\label{h11}
\partial_t(\rho \mathbf{u})+\bm{\nabla} \cdot\left(\rho \mathbf{u} \mathbf{u}+P \mathbf{I}+\bm{\Delta}_2^*\right)=0,
\end{equation}
\begin{equation}\label{h12}
\partial_t(\rho e)+\bm{\nabla} \cdot\left[(\rho e+P) \mathbf{u}+\bm{\Delta}_2^* \cdot \mathbf{u}+\bm{\Delta}_{3,1}^*\right]=0.
\end{equation}
Here, $P=\rho R T$ is the pressure, $e=c_v T+u^2 / 2$ is the specific total energy, and $c_v=(n+2) R / 2$ is the specific heat at constant volume. $\bm{\Delta}_2^*$ and $\bm{\Delta}_{3,1}^*$ denote the non-organized momentum flux (NOMF) and non-organized energy flux (NOEF), respectively.

At first order in the CE expansion, the constitutive relations reduce to the NS forms:
\begin{equation}\label{h13}
\bm{\Delta}_2^{*(1)}=-\mu\left[\bm{\nabla} \mathbf{u}+(\bm{\nabla} \mathbf{u})^T-\frac{2}{n+2} \mathrm{I} \bm{\nabla} \cdot \mathbf{u}\right],
\end{equation}
\begin{equation}\label{h14}
\bm{\Delta}_{3,1}^{*(1)}=-\kappa \bm{\nabla} T.
\end{equation}
The corresponding viscosity and thermal conductivity are
\begin{equation}\label{h15}
\mu(\rho, T)=P \tau(\rho, T),
\end{equation}
\begin{equation}\label{h16}
\kappa(\rho, T)=c_p P \tau(\rho, T),
\end{equation}
where $c_p=(n+4) R / 2$ is the specific heat at constant pressure. Therefore, once $\tau=\tau(\rho,T)$ is introduced, the transport coefficients inherit the same state dependence.

At second order in the CE expansion, Burnett-level corrections arise and account for higher-order nonequilibrium transport. In the present DTRT-DBM, the analytical expressions for the second-order components of the non-organized momentum flux $\bm{\Delta}_2^*$ and non-organized energy flux $\bm{\Delta}_{3,1}^*$ are given in appendix~\ref{B}.

Although CE analysis provides a convenient test of hydrodynamic recoverability, it should be emphasized that DBM is not defined by the CE procedure alone. In DBM, the physical description is ultimately provided by the extended hydrodynamic equations (EHEs), which include not only the evolution equations for conserved quantities but also the explicit dynamics of selected non-conserved quantities, thereby directly characterizing the departure of the system from local equilibrium \citep{xu2012FOP, xu2018KT, Xu2022BSTP, xu2024FOP}.

In complex systems, nonlinearity and nonequilibrium are often much more pronounced than can be adequately represented by macroscopic indicators such as the Knudsen number, spatial gradients, or other continuum-level measures alone. DBM retains these traditional indicators, but further constructs a systematic framework for describing nonequilibrium through higher-order non-conserved kinetic moments:
\begin{equation}\label{h17}
\begin{aligned}
\bm{\Delta}_{m, n}^* & =\mathbf{M}_{m, n}^*\left(f-f^{eq}\right) \\
& =\sum_i\left(\frac{1}{2}\right)^{1-\delta_{m, n}}\left(f_i-f_i^{eq}\right) \underbrace{\mathbf{v}_i^* \mathbf{v}_i^* \cdots \mathbf{v}_i^*}_n\left(\mathbf{v}_i^{* 2}+\eta_i^2\right)^{\frac{m-n}{2}}.
\end{aligned}
\end{equation}
Here, $m$ denotes the tensor order before contraction, and $n$ denotes the order after contraction, while $\mathbf{v}_i^*=\mathbf{v}_i-\mathbf{u}$ is the particle peculiar velocity relative to the macroscopic flow velocity. The corresponding kinetic moments $\mathbf{M}_{m,n}^*$ are central moments. Through the nonequilibrium measures $\bm{\Delta}_{m,n}^*$, DBM can characterize both hydrodynamic nonequilibrium effects (HNEs) and thermodynamic nonequilibrium effects (TNEs), thereby providing a mesoscopic perspective on departures from equilibrium.

Within this framework, we further introduce a multidimensional indicator vector to summarize the nonequilibrium state from multiple perspectives:
\begin{equation}\label{h18}
S_{TNE}=\left\{\bm{\Delta}^*,|\bm{\Delta}|,|\bm{\nabla}\mathbf{u}|,|\bm{\nabla}\rho|,|\bm{\nabla}T|,\tau_0,a,b,Kn,\cdots\right\}.
\end{equation}
This indicator system incorporates local nonequilibrium measures, global nonequilibrium intensity, macroscopic gradient features, and model parameters into a unified framework, thus providing a mesoscopic basis for nonequilibrium analysis and feature identification in complex flows \citep{xu2018KT, zhang2022POF}.

\section{Model Validation}\label{III}

This section validates the DTRT-DBM at both the macroscopic and mesoscopic levels. We first examine its ability to recover the canonical wave structures of the Sod shock tube problem, namely the shock wave, contact discontinuity, and rarefaction wave. We then assess its accuracy in describing nonequilibrium quantities by comparing numerical and analytical results for viscous stress and heat flux.

\subsection{Numerical framework and discrete velocity model}\label{III-0}

As a theoretical framework for model construction and for analysing complex physical fields, DBM specifies only the essential physical constraints of the problem and does not prescribe particular discretization schemes in time, space, or velocity. In the present work, spatial derivatives are computed with a fifth-order weighted essentially non-oscillatory (WENO) finite-difference scheme to suppress spurious oscillations near steep gradients \citep{jiang1996JCP}, while temporal integration is performed with a second-order implicit--explicit Runge--Kutta scheme \citep{ascher1997ANM}.

\begin{figure}\small
	\centering
	\includegraphics[width=0.55\textwidth]{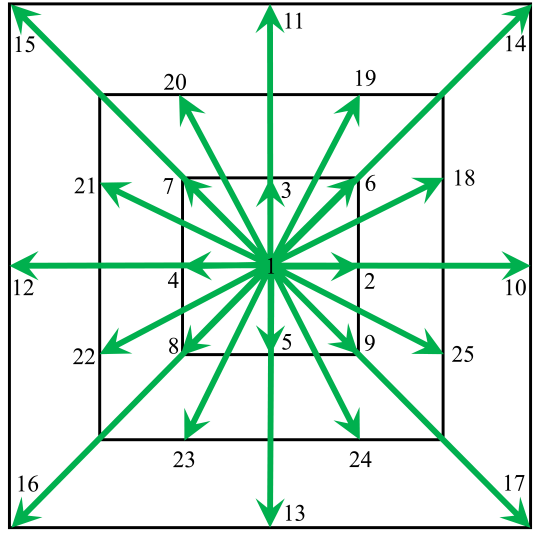}
	\caption{Discrete velocity set of the D2V25 model.}
	\label{Fig01}
\end{figure}

The accuracy and efficiency of DBM also depend critically on phase-space discretization. In particular, the discrete velocity set (DVS) should satisfy the requirements of symmetry, full rank, and sufficient diversity in both speed and direction \citep{he2025POF}. To balance accuracy and stability, all simulations in this work employ the two-dimensional 25-velocity model (D2V25), shown in figure~\ref{Fig01}. The D2V25 discrete velocity set is defined as follows \citep{he2025POF}:
\begin{equation}\label{h19}
\left(v_{i x}, v_{i y}\right)=\left\{\begin{array}{c}
c(0,0), i=1 \\
cyc: c( \pm 1,0), 2 \leq i \leq 5 \\
c( \pm 1, \pm 1), 6 \leq i \leq 9 \\
cyc: c( \pm 3,0), 10 \leq i \leq 13 \\
c( \pm 3, \pm 3), 14 \leq i \leq 17 \\
cyc: c( \pm 2, \pm 1), 18 \leq i \leq 25.
\end{array}\right.
\end{equation}

%%%%%%%%%%%%%%%%%%%%%%%%%%%%%%%

The auxiliary variable is specified as follows: $\eta_i=4\eta_0$ for $1\leq i\leq2$, $\eta_i=3\eta_0$ for $i=3$, $\eta_i=2\eta_0$ for $i=4$, $\eta_i=\eta_0$ for $i=5$ and $14\leq i\leq17$, and $\eta_i=0$ otherwise. Here, ``cyc'' denotes cyclic permutation, and $c$ and $\eta_0$ are free parameters. With suitable choices of these parameters, the matrix $\mathrm{C}$ remains invertible, and numerical stability and efficiency are improved.

\subsection{Macroscopic validation: one-dimensional Sod shock tube}\label{III-1}

The Sod shock tube problem is a standard benchmark for assessing a model's ability to capture strong discontinuities and multi-wave propagation. It contains the three canonical wave structures of a shock wave, a contact discontinuity, and a rarefaction wave, and is therefore well suited to validating the macroscopic accuracy of the present model. The initial conditions are
\begin{equation}\label{h20}
\left\{\begin{array}{l}
\left.\left(\rho, T, u_x, u_y\right)\right|_L=(1.0,1.0,0.0,0.0), \\
\left.\left(\rho, T, u_x, u_y\right)\right|_R=(0.125,0.8,0.0,0.0).
\end{array}\right.
\end{equation}
Here, the subscripts ``$L$'' and ``$R$'' denote the left and right states, respectively. The computational grid is $N_x \times N_y=1000 \times 4$, with $\Delta x=\Delta y=10^{-3}$ and $\Delta t=10^{-4}$. The remaining model parameters are $\tau_0=5 \times 10^{-5}$, $\rho_0=T_0=1.0$, $c=1.05$, and $\eta_0=1.0$.

Periodic boundary conditions are imposed in the $y$-direction. At the left boundary in the $x$-direction, we prescribe
\begin{equation}\label{h21}
f_{i, -1, t}=f_{i, 0, t}=f_{i, 1, t=0}^{(0)},
\end{equation}
which implies
\begin{equation}\label{h22}
\left(\rho, T, u_x, u_y\right)_{-1, t}=\left(\rho, T, u_x, u_y\right)_{0, t}=\left(\rho, T, u_x, u_y\right)_{1, t=0}.
\end{equation}
The same equilibrium boundary treatment is applied at the right boundary:
\begin{equation}\label{h23}
f_{i, N_x+2, t}=f_{i, N_x+1, t}=f_{i, N_x, t=0}^{(0)},
\end{equation}
\begin{equation}\label{h24}
\left(\rho, T, u_x, u_y\right)_{N_x+2, t}=\left(\rho, T, u_x, u_y\right)_{N_x+1, t}=\left(\rho, T, u_x, u_y\right)_{N_x, t=0}.
\end{equation}
The relaxation time is prescribed by equation~(\ref{h3}), where $a$ and $b$ determine its dependence on the local macroscopic state.

\subsubsection{Weak-dissipation case: agreement with the Riemann solution}\label{III-1-1}

%%%%%%%%%%%%%%%%%%%%%%%%%%%%%%%
\begin{figure}\small
	\centering
	\includegraphics[width=0.82\textwidth]{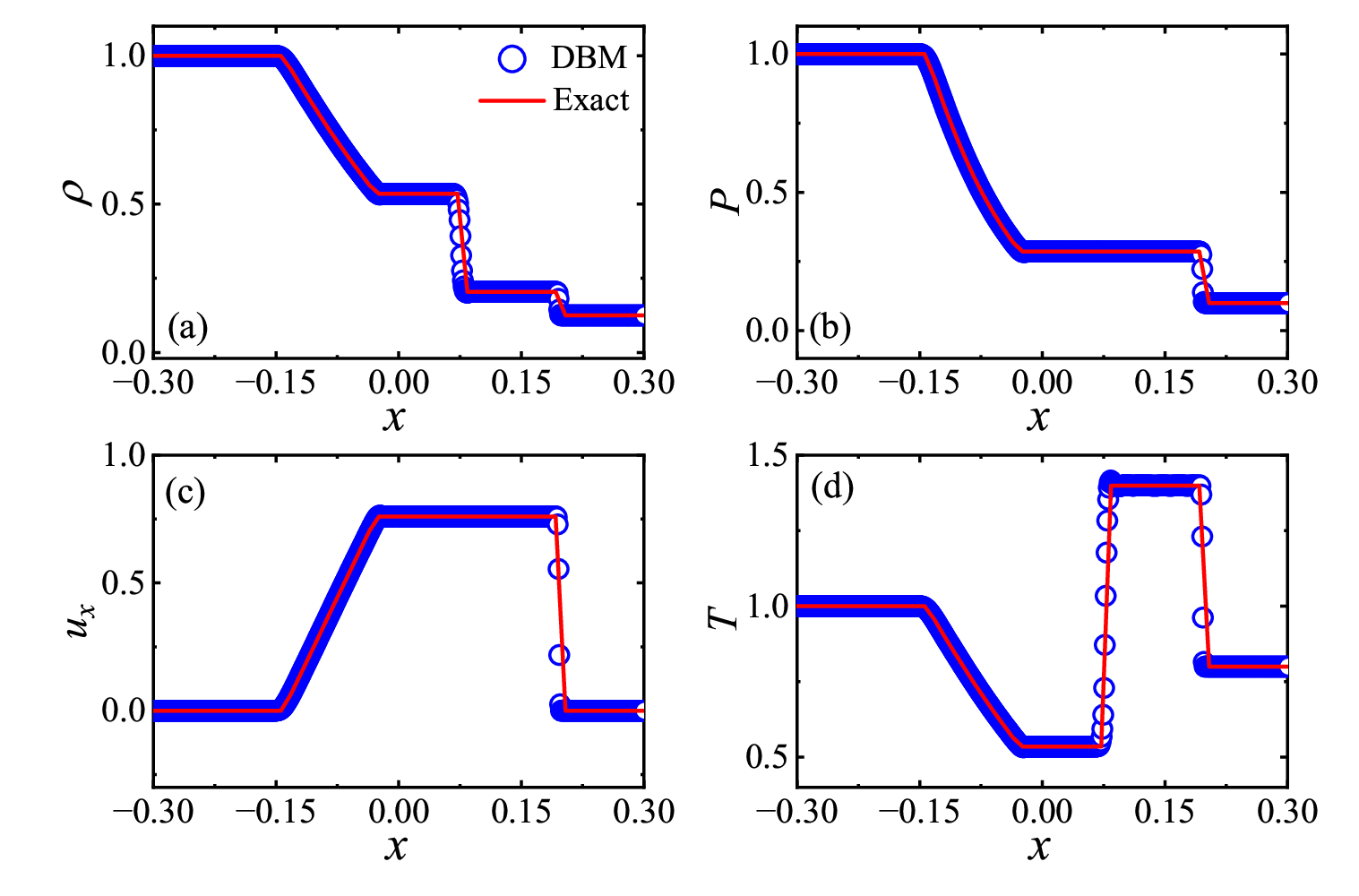}
	\caption{Comparison of the DTRT-DBM results with the exact Riemann solution for the Sod shock tube at $t=0.1$ with $a=10$ and $b=10$: (a) density, (b) pressure, (c) velocity, and (d) temperature.}
	\label{Fig02}
\end{figure}

Figure~\ref{Fig02} compares the DTRT-DBM results at $t=0.1$ with the exact Riemann solution for $a=b=10$. The numerical profiles of density, pressure, velocity, and temperature agree closely with the exact solution. The rarefaction wave, contact discontinuity, and shock wave are all captured accurately without spurious oscillations. For this case, the relaxation time is
\[
\tau=\tau_0\left(\frac{\rho}{\rho_0}\right)^a\left(\frac{T}{T_0}\right)^b=\tau_0 \rho^{10} T^{10}=\tau_0 P^{10}.
\]
Figure~\ref{Fig02}(b) shows that the pressure remains below $1.0$ throughout the computational domain. Hence, $\tau \ll \tau_0$. The nonequilibrium effect is therefore weak, and the flow remains close to the Euler limit.

\subsubsection{Strong-dissipation case: enhanced nonequilibrium effects}\label{III-1-2}

%%%%%%%%%%%%%%%%%%%%%%%%%%%%%%%
\begin{figure}\small
	\centering
	\includegraphics[width=0.82\textwidth]{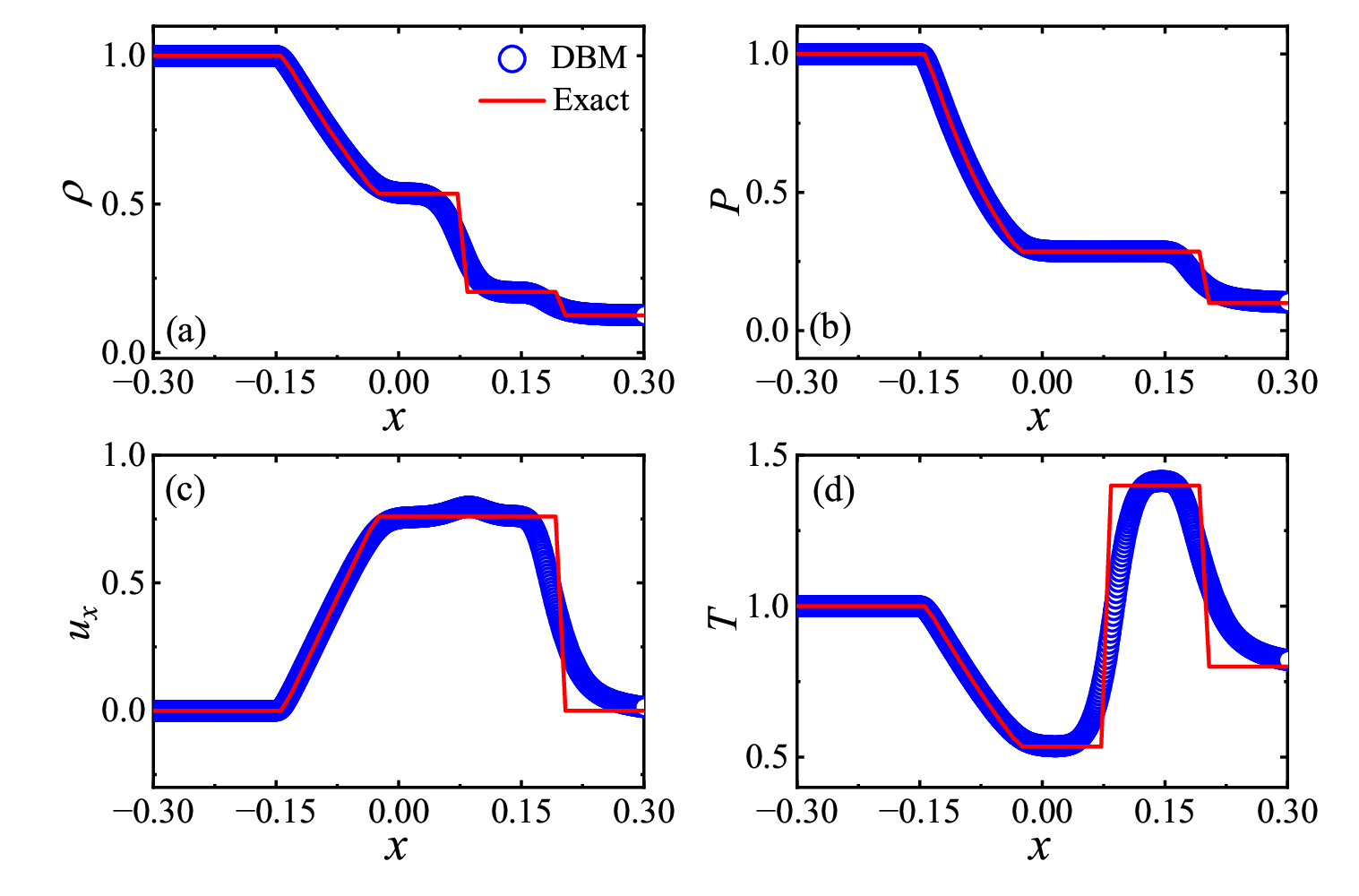}
	\caption{Comparison of the DTRT-DBM results with the exact Riemann solution for the Sod shock tube at $t=0.1$ with $a=-3$ and $b=-3$: (a) density, (b) pressure, (c) velocity, and (d) temperature.}
	\label{Fig03}
\end{figure}
%%%%%%%%%%%%%%%%%%%%%%%%%%%%%%%
\begin{figure}\small
	\centering
	\includegraphics[width=0.82\textwidth]{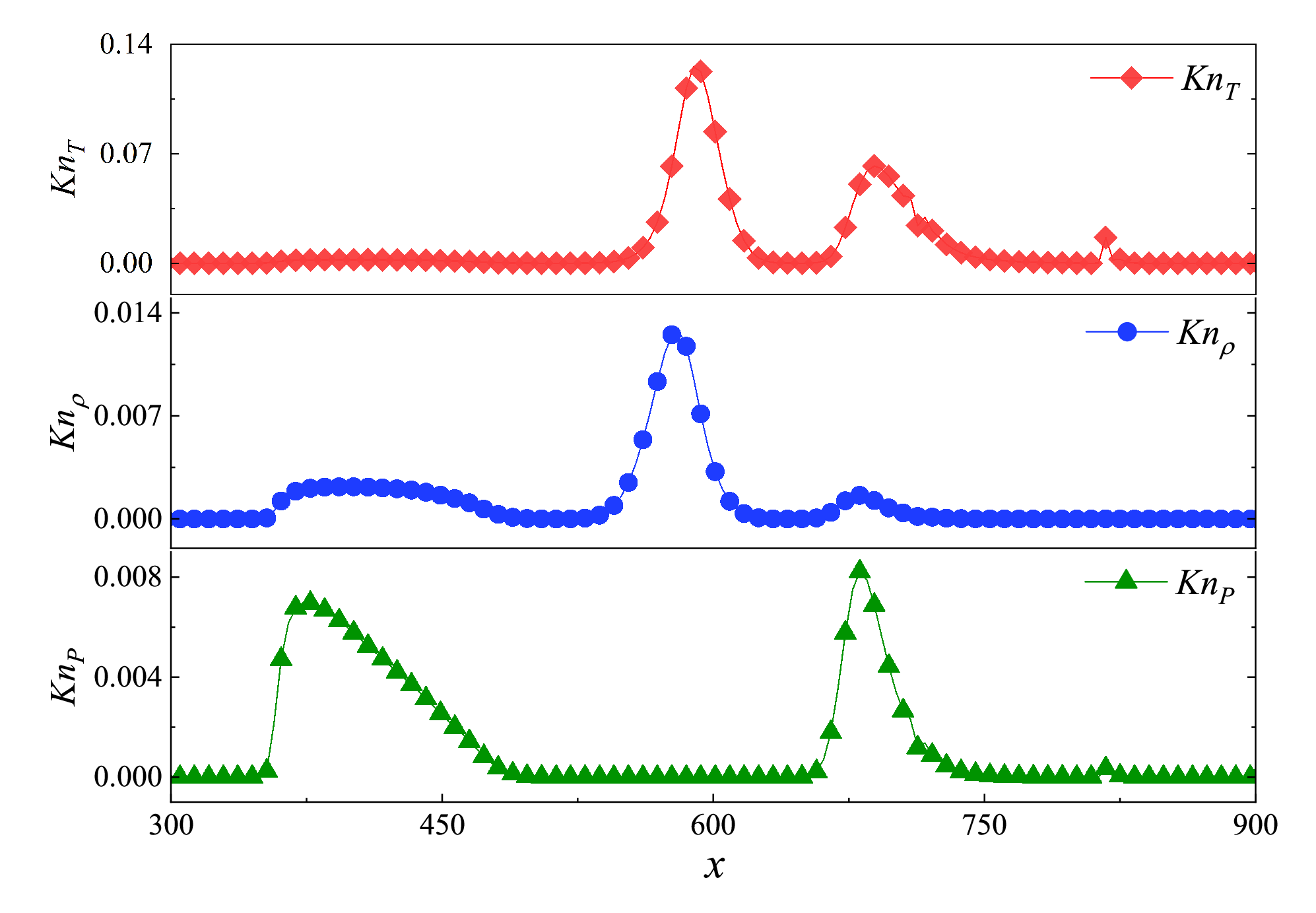}
	\caption{Profile of the local Knudsen numbers calculated from $T$, $\rho$ and $P$ along the line $y=0.5L_y$ at $t=0.1$ for $a=-3$ and $b=-3$.}
	\label{Fig04}
\end{figure}

Figure~\ref{Fig03} compares the numerical macroscopic profiles at $t=0.1$ with the exact Riemann solution for $a=b=-3$. The density and pressure deviate from the exact Riemann solution near the discontinuities, as shown in figures~\ref{Fig03}(a,b). More pronounced deviations appear in the velocity and temperature profiles in figures~\ref{Fig03}(c,d), where both the shock front and the contact discontinuity are visibly broadened. The corresponding relaxation time is
\[
\tau=\tau_0 P^{-3}.
\]

In the initial right state, $P=0.1$, and hence $\tau=10^3 \tau_0$. In low-pressure regions, the local state causes $\tau$ to increase substantially, thereby enhancing nonequilibrium transport and smoothing the flow field. By contrast, in high-pressure regions, $\tau$ decreases dynamically, so that local dissipation weakens and steep gradients are better preserved. This spatial variation in the relaxation time produces different broadening widths and displacement magnitudes for the rarefaction wave, contact discontinuity, and shock wave.

Figure~\ref{Fig04} shows the corresponding profile of the local $Kn$ along the line $y=0.5L_y$. The local $Kn$ reaches the order of $0.1$, indicating that the flow lies outside the range of validity of the Euler equations. This test shows that the DTRT-DBM can represent state-dependent dissipative effects in strongly nonequilibrium regimes.

\subsection{Mesoscopic validation: nonequilibrium quantities}\label{III-2}

We next validate the nonequilibrium description of the DTRT-DBM by comparing numerical and analytical solutions for viscous stress and heat flux. These tests assess the mesoscopic accuracy of the model and examine whether the state-dependent relaxation time can correctly represent the corresponding nonequilibrium transport.

\subsubsection{Viscous stress}\label{III-2-1}

To verify the accuracy of the DTRT-DBM in describing viscous stress, we prescribe the initial conditions as
\begin{equation}\label{h27}
\rho(x, y)=\frac{\rho_L+\rho_R}{2}-\frac{\rho_L-\rho_R}{2} \tanh \left(\frac{x-(N_x \Delta x) / 2}{L_\rho}\right),
\end{equation}
\begin{equation}\label{h28}
u_x(x, y)=-u_0 \tanh \left(\frac{x-(N_x \Delta x) / 2}{L_u}\right), \quad u_y=0.
\end{equation}
Here, $\rho_L=1$, $\rho_R=2$, $T_L=T_R=1$, and $L_\rho=L_u=20$. The remaining parameters are $\Delta x=\Delta y=0.002$, $\Delta t=5 \times 10^{-5}$, $c=1.05$, $\eta_0=1.0$, $\gamma=2$, $u_0=0.5$, $\tau_0=5 \times 10^{-4}$, and $\rho_0=T_0=1$.

%%%%%%%%%%%%%%%%%%%%%%%%%%%%%%%
\begin{figure}\small
	\centering
	\includegraphics[width=0.98\textwidth]{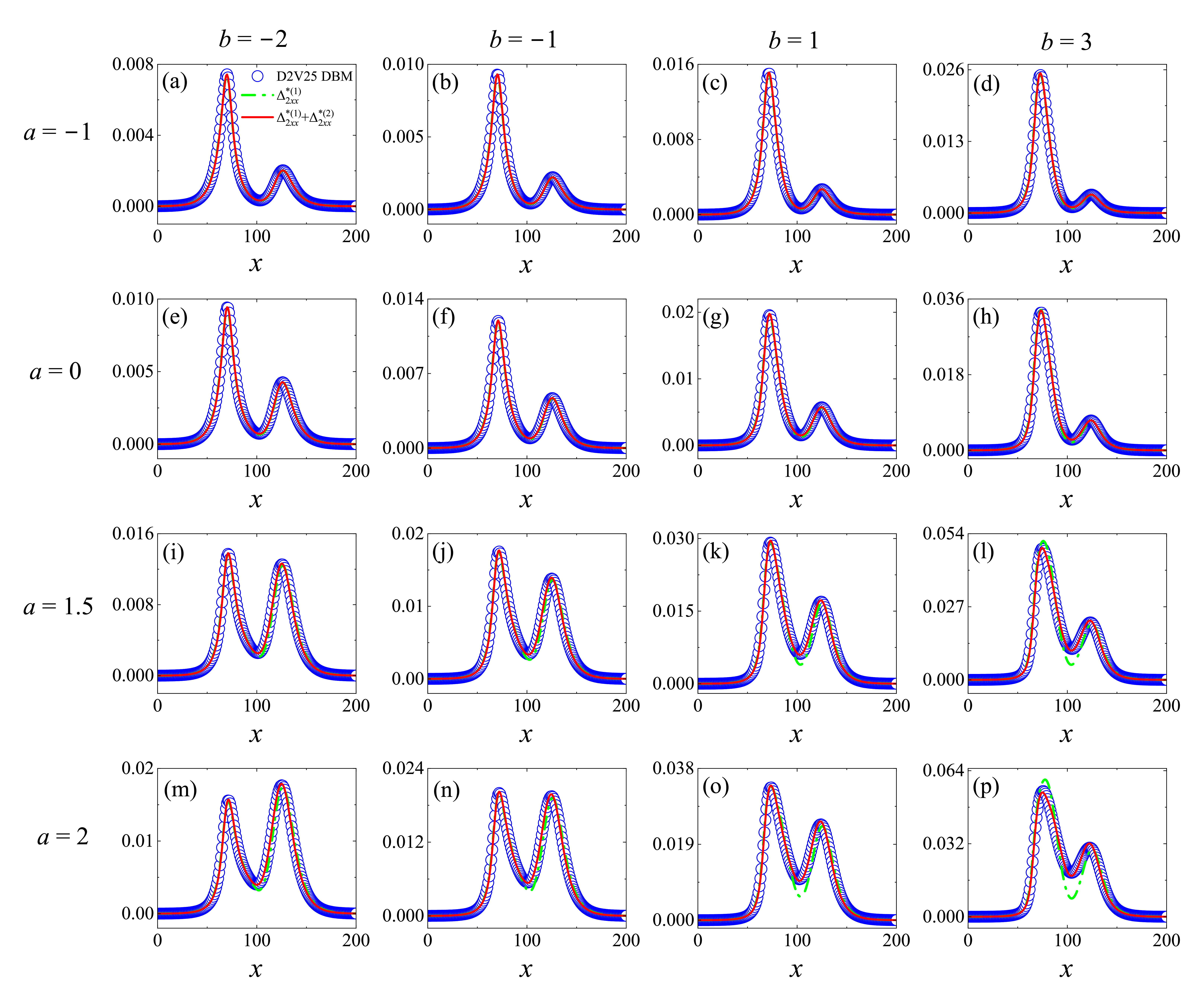}
	\caption{Profiles of viscous stress along the centreline $y=0.5L_y$ at $t=0.04$ for different parameter combinations $(a=-1,0,1.5,2;\, b=-2,-1,1,3)$. DTRT-DBM results are compared with the corresponding analytical solutions.}
	\label{Fig05}
\end{figure}

Figure~\ref{Fig05} shows viscous-stress profiles along the centreline at $t=0.04$ for different parameter combinations $(a=-1,0,1.5,2;\, b=-2,-1,1,3)$, together with the corresponding analytical solutions. In all cases, the numerical results agree well with the second-order analytical solutions. This agreement holds not only in weakly nonequilibrium states but also when second-order nonequilibrium effects become significant. These results confirm that the DTRT-DBM remains accurate over a wide range of nonequilibrium conditions.

\paragraph{(i) Dependence of the interfacial nonequilibrium structure on $a$ and $b$.}

As shown in figure~\ref{Fig05}, the viscous stress exhibits a typical double-peak structure. The left peak, located approximately at $x=50\sim100$, is higher, whereas the right peak, located approximately at $x=100\sim150$, is lower. This asymmetry arises because the parameters $a$ and $b$ modify the spatial distribution of $\tau$. In the present case, the density is higher on the right-hand side. As $a$ increases, $\tau$ rises more strongly on the right, and the right peak is enhanced accordingly. By contrast, increasing $b$ affects the left peak more strongly and further enlarges the difference between the two peaks. Therefore, when $a$ is relatively large and $b$ is relatively small, the two peaks approach each other; when $a$ is relatively small and $b$ is relatively large, the right peak weakens and the interfacial asymmetry becomes more pronounced. Thus, the observed interfacial structure reflects how the local relaxation time varies with the model parameters.

To quantify the difference between the two peaks, we introduce the logarithmic asymmetry index
\begin{equation}\label{h29}
D=\ln \left|\frac{\Delta L}{\Delta R}\right|.
\end{equation}
Here, $\Delta L$ and $\Delta R$ denote the left and right peak values, respectively. The condition $D=0$ indicates that the two peaks are symmetric in magnitude; $D>0$ indicates that the left peak is larger; and $D<0$ indicates that the right peak is larger. A larger $|D|$ corresponds to a stronger asymmetry.

%%%%%%%%%%%%%%%%%%%%%%%%%%%%%%%%%%%%%%%%%%%%%%%%%%%%%%%%%%%%%%%%%%%%
\begin{center}
\captionof{table}{Values of the logarithmic asymmetry index $D$ for the double-peak viscous-stress structure under different parameter combinations.}
\label{tab01}

\renewcommand{\arraystretch}{1.2}
\setlength{\tabcolsep}{7pt}

\begin{tabular}{>{\centering\arraybackslash}p{2.0cm}
                >{\centering\arraybackslash}p{2.15cm}
                >{\centering\arraybackslash}p{2.15cm}
                >{\centering\arraybackslash}p{2.15cm}
                >{\centering\arraybackslash}p{2.15cm}}
\hline
      & $b=-2$ & $b=-1$ & $b=1$ & $b=3$ \\
\hline
$a=-1$   & $1.31$  & $1.43$  & $1.71$  & $2.00$ \\
$a=0$    & $0.80$  & $0.93$  & $1.23$  & $1.54$ \\
$a=1.5$  & $0.86$  & $0.24$  & $0.55$  & $0.80$ \\
$a=2$    & $-0.13$ & $0.02$  & $0.31$  & $0.53$ \\
\hline
\end{tabular}
\end{center}
%%%%%%%%%%%%%%%%%%%%%%%%%%%%%%%%%%%%%%%%%%%%%%%%%%%%%%%%%%%%%%%%%%%%

%%%%%%%%%%%%%%%%%%%%%%%%%%%%%%%
\begin{figure}\small
	\centering
	\includegraphics[width=0.82\textwidth]{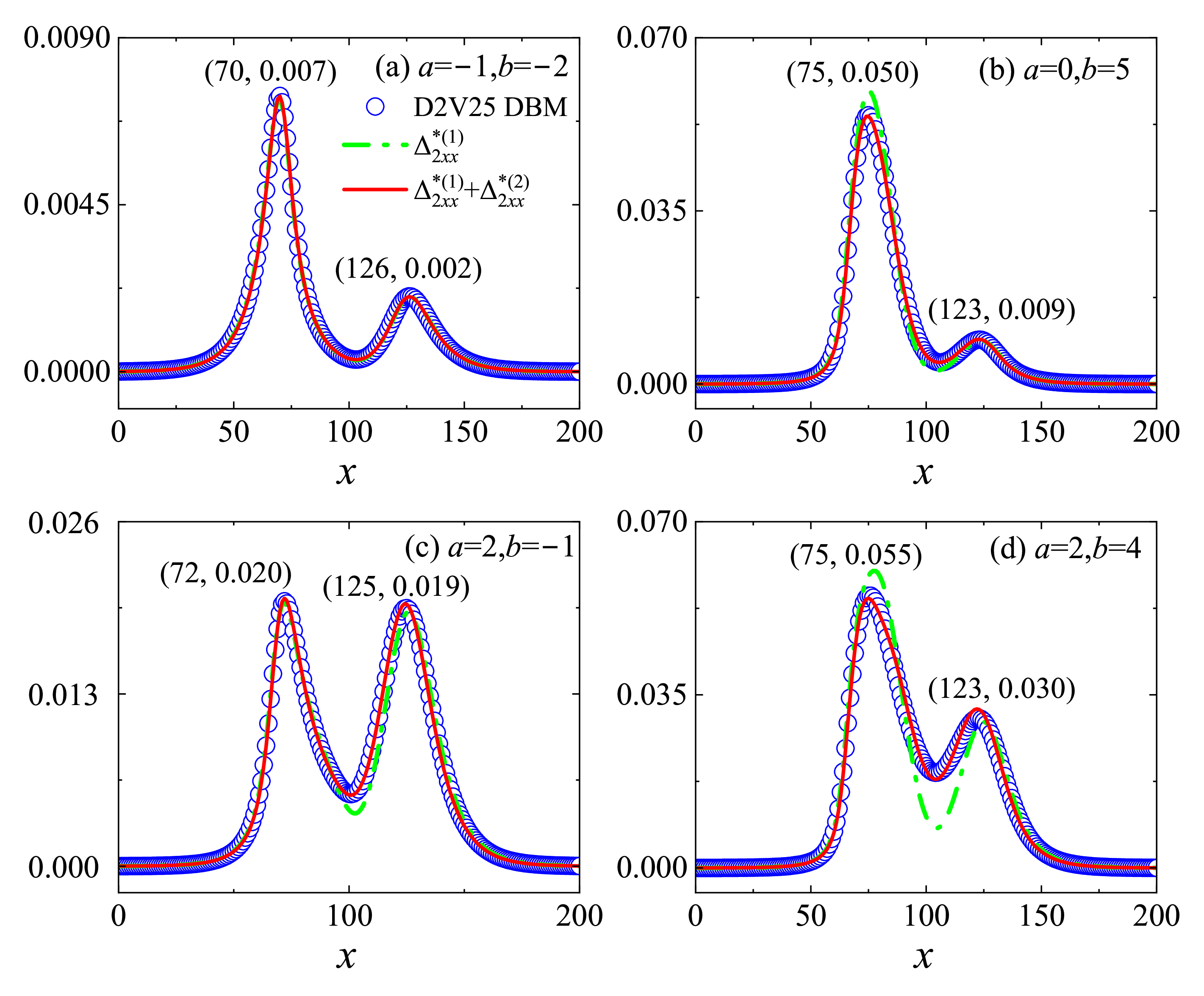}
	\caption{Profiles of $\Delta_{2xx}^{*}$ along the centreline $y=0.5L_y$ at $t=0.04$ for four representative cases: (a) $a=-1,\, b=-2$; (b) $a=0,\, b=5$; (c) $a=2,\, b=-1$; and (d) $a=2,\, b=4$.}
	\label{Fig06}
\end{figure}

To clarify how the nonequilibrium intensity varies with $a$ and $b$ at different orders, we introduce the relative nonequilibrium intensity index $R_{\text{TNE}}$. Figure~\ref{Fig06} shows four representative viscous-stress profiles, and table~\ref{tab02} lists the corresponding $R_{\text{TNE}}$ values and nonequilibrium states.

%%%%%%%%%%%%%%%%%%%%%%%%%%%%%%%%%%%%%%%%%%%%%%%%%%%%%%%%%%%%%%%%%%%%
\begin{center}
\captionof{table}{Relative nonequilibrium intensity $R_{\text{TNE}}$ and the corresponding nonequilibrium states for viscous stress under different parameter combinations.}
\label{tab02}

\renewcommand{\arraystretch}{1.2}
\setlength{\tabcolsep}{7pt}

\begin{tabular}{>{\centering\arraybackslash}p{3.8cm}
                >{\centering\arraybackslash}p{2.2cm}
                >{\centering\arraybackslash}p{5.6cm}}
\hline
Parameter Combination & $R_{\text{TNE}}$ & Relative Nonequilibrium State in the System \\
\hline
$a=-1,\; b=-2$ & $0$    & Zero Relative Nonequilibrium \\
$a=0,\; b=5$   & $0.15$ & Weak Relative Nonequilibrium \\
$a=2,\; b=-1$  & $0.5$  & Moderate Relative Nonequilibrium \\
$a=2,\; b=4$   & $1.1$  & Strong Relative Nonequilibrium \\
\hline
\end{tabular}
\end{center}
%%%%%%%%%%%%%%%%%%%%%%%%%%%%%%%%%%%%%%%%%%%%%%%%%%%%%%%%%%%%%%%%%%%%

Figure~\ref{Fig06} and table~\ref{tab02} show that, when both $a$ and $b$ are small, the overall viscous-stress intensity remains low and first-order nonequilibrium effects dominate ($R_{TNE}=0$). When one parameter is large and the other is small, their effects on the relative nonequilibrium intensity become competitive: the smaller parameter suppresses the nonequilibrium level, whereas the larger one enhances the local transport response. The system then lies in a weak-to-moderate relative nonequilibrium state. When both $a$ and $b$ increase simultaneously, their combined effect becomes significant. The relaxation rate becomes strongly non-uniform in space, leading to a marked increase in viscous-stress nonequilibrium. In this case, $R_{TNE}$ reaches $1.1$, corresponding to a strong relative nonequilibrium state. These results show how different parameter combinations are reflected in the nonequilibrium state of the system.

\subsubsection{Heat flux}\label{III-2-2}

\begin{figure}\small
	\centering
	\includegraphics[width=0.98\textwidth]{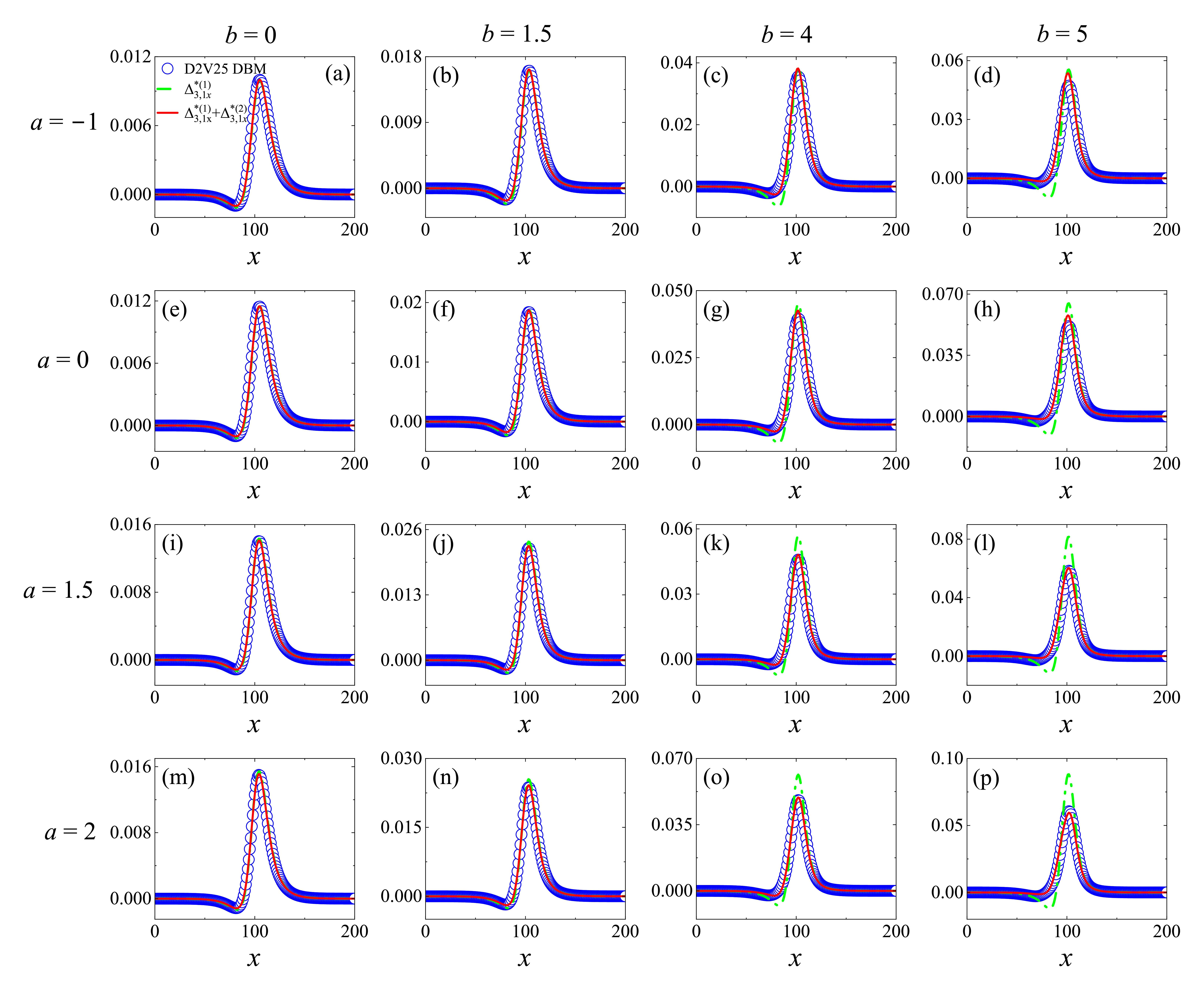}
	\caption{Profiles of heat flux along the centreline $y=0.5L_y$ at $t=0.0075$ for different parameter combinations $(a=-1,0,1.5,2;\, b=0,1.5,4,5)$. DTRT-DBM results are compared with the corresponding analytical solutions.}
	\label{Fig07}
\end{figure}

To verify the accuracy of the model in describing heat flux, we use the same initial conditions as in section~\ref{III-2-1}, but with $\rho_L=\rho_R=1$, $T_L=1.5$, $T_R=1$, $\Delta x=\Delta y=0.003$, and $u_0=1.2$. Figure~\ref{Fig07} shows heat flux profiles along the centreline at $t=0.0075$ for 16 parameter combinations $(a=-1,0,1.5,2;\, b=0,1.5,4,5)$. In all cases, the numerical and analytical solutions agree well, confirming that the model accurately captures heat flux nonequilibrium.

%%%%%%%%%%%%%%%%%%%%%%%%%%%%%%%

\paragraph{(i) Dependence of the interfacial nonequilibrium structure on $a$ and $b$.}

The heat flux also exhibits a double-peak structure. The right peak, located approximately at $x=100\sim150$, is positive and larger in magnitude, whereas the left peak, located approximately at $x=50\sim100$, is negative and smaller in magnitude. The corresponding values of $D$ are listed in table~\ref{tab03}.

%%%%%%%%%%%%%%%%%%%%%%%%%%%%%%%%%%%%%%%%%%%%%%%%%%%%%%%%%%%%%%%%%%%%
\begin{center}
\captionof{table}{Values of the logarithmic asymmetry index $D$ for the double-peak heat flux structure under different parameter combinations.}
\label{tab03}

\renewcommand{\arraystretch}{1.2}
\setlength{\tabcolsep}{7pt}

\begin{tabular}{>{\centering\arraybackslash}p{2.0cm}
                >{\centering\arraybackslash}p{2.15cm}
                >{\centering\arraybackslash}p{2.15cm}
                >{\centering\arraybackslash}p{2.15cm}
                >{\centering\arraybackslash}p{2.15cm}}
\hline
      & $b=0$ & $b=1.5$ & $b=4$ & $b=5$ \\
\hline
$a=-1$   & $-2.29$ & $-2.31$ & $-2.69$ & $-3.50$ \\
$a=0$    & $-2.40$ & $-2.36$ & $-2.78$ & $-3.82$ \\
$a=1.5$  & $-2.49$ & $-2.53$ & $-2.90$ & $-3.91$ \\
$a=2$    & $-2.54$ & $-2.55$ & $-2.92$ & $-4.07$ \\
\hline
\end{tabular}
\end{center}
%%%%%%%%%%%%%%%%%%%%%%%%%%%%%%%%%%%%%%%%%%%%%%%%%%%%%%%%%%%%%%%%%%%%

Table~\ref{tab03} shows that the heat flux asymmetry increases continuously as the parameters vary. All values of $D$ are negative, indicating that the magnitude of the left peak remains smaller than that of the right peak. As $a$ increases, the right peak is slightly enhanced, whereas the left peak changes only weakly. As $b$ increases, the right peak is enhanced much more strongly, and the asymmetry between the two peaks is amplified markedly. When both $a$ and $b$ are large, their combined effect is strongest: the right peak rises markedly, the left peak weakens further, and $|D|$ reaches a maximum of $4.07$.

This behaviour is consistent with the initial temperature distribution. Since $T_L=1.5$ and $T_R=1.0$, heat is transported from left to right. The right peak therefore attains a larger magnitude, whereas the left peak remains negative. The parameter $b$ increases the thermal conductivity more strongly and thereby makes the right peak significantly larger than the left peak.

\paragraph{(ii) Dependence of the relative nonequilibrium intensity on $a$ and $b$.}

%%%%%%%%%%%%%%%%%%%%%%%%%%%%%%%
\begin{figure}\small
	\centering
	\includegraphics[width=0.82\textwidth]{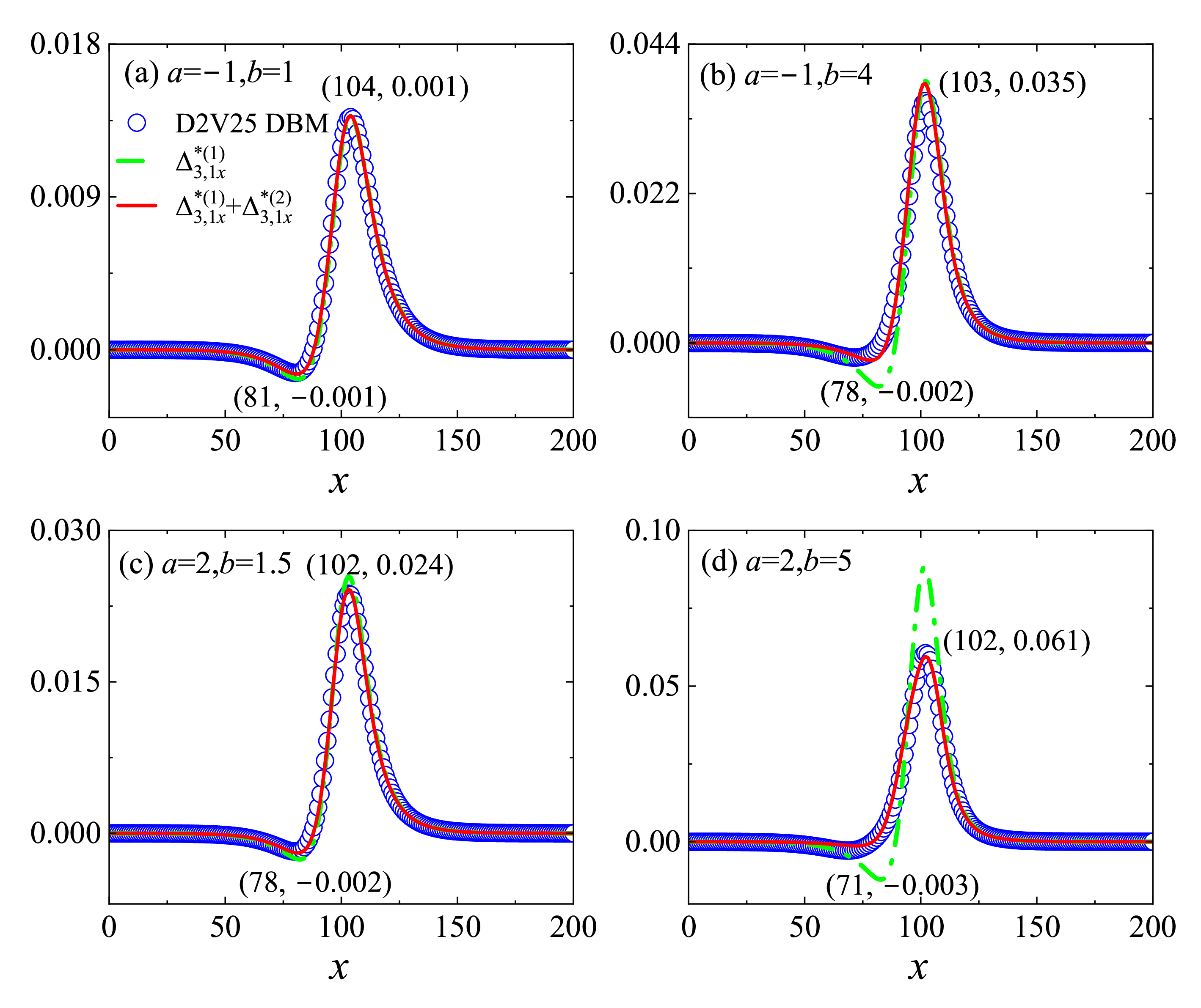}
	\caption{Profiles of $\Delta_{3,1x}^{*}$ along the centreline $y=0.5L_y$ at $t=0.0075$ for four representative cases: (a) $a=-1,\, b=1$; (b) $a=-1,\, b=4$; (c) $a=2,\, b=1.5$; and (d) $a=2,\, b=5$.}
	\label{Fig08}
\end{figure}

%%%%%%%%%%%%%%%%%%%%%%%%%%%%%%%%%%%%%%%%%%%%%%%%%%%%%%%%%%%%%%%%%%%%
\begin{center}
\captionof{table}{Relative nonequilibrium intensity $R_{\text{TNE}}$ and the corresponding nonequilibrium states for heat flux under different parameter combinations.}
\label{tab04}

\renewcommand{\arraystretch}{1.2}
\setlength{\tabcolsep}{7pt}

\begin{tabular}{>{\centering\arraybackslash}p{3.8cm}
                >{\centering\arraybackslash}p{2.2cm}
                >{\centering\arraybackslash}p{5.6cm}}
\hline
Parameter Combination & $R_{TNE}$ & Relative Nonequilibrium State in the System \\
\hline
$a=-1,\; b=1$   & $0$   & Zero Relative Nonequilibrium \\
$a=-1,\; b=4$   & $0.6$ & Moderate Relative Nonequilibrium \\
$a=2,\; b=1.5$  & $0.2$ & Weak Relative Nonequilibrium \\
$a=2,\; b=5$    & $1.1$ & Strong Relative Nonequilibrium \\
\hline
\end{tabular}
\end{center}
%%%%%%%%%%%%%%%%%%%%%%%%%%%%%%%%%%%%%%%%%%%%%%%%%%%%%%%%%%%%%%%%%%%%

To clarify the effects of $a$ and $b$ on the relative nonequilibrium intensity of heat flux, figure~\ref{Fig08} shows four representative distributions, and table~\ref{tab04} lists the corresponding $R_{TNE}$ values. These results show that, when both $a$ and $b$ are small $(a=-1,b=1)$, the overall heat flux nonequilibrium remains weak, first-order nonequilibrium effects dominate, and $R_{\text{TNE}}\approx 0$. When one parameter is large and the other is small, as in $(a=-1, b=4)$ or $(a=2, b=1.5)$, their effects on $R_{\text{TNE}}$ become competitive: the larger parameter enhances nonequilibrium, whereas the smaller one suppresses it. The system then remains in a weak-to-moderate nonequilibrium state, with the effect of $b$ being more pronounced. When both $a$ and $b$ increase simultaneously, as in $(a=2, b=5)$, their combined effect becomes significant, the heat flux magnitude increases, and $R_{\text{TNE}}$ reaches $1.1$, corresponding to a strong relative nonequilibrium state.

Overall, the DTRT-DBM provides accurate descriptions of both macroscopic flow evolution and nonequilibrium effects. The results in this section show that the model can capture both the nonequilibrium measures and the relative nonequilibrium intensity of the flow over a wide range of parameter combinations. The physical mechanisms underlying these behaviours are analysed further through nonequilibrium phase diagrams in the next section.

\section{Nonequilibrium phase diagrams}\label{IV}

This section examines how the density exponent $a$ and the temperature exponent $b$ influence the nonequilibrium characteristics through the state-dependent relaxation time. The analysis is based on phase diagrams of the extrema of two representative nonequilibrium quantities, viscous stress and heat flux, together with the corresponding variations of the fitted slopes with the relaxation parameters.

\subsection{Viscous stress}\label{IV-1}

\subsubsection{Phase diagram of the total viscous stress}\label{IV-1-1}

%%%%%%%%%%%%%%%%%%%%%%%%%%%%%%%
\begin{figure}\small
	\centering
	\includegraphics[width=0.65\textwidth]{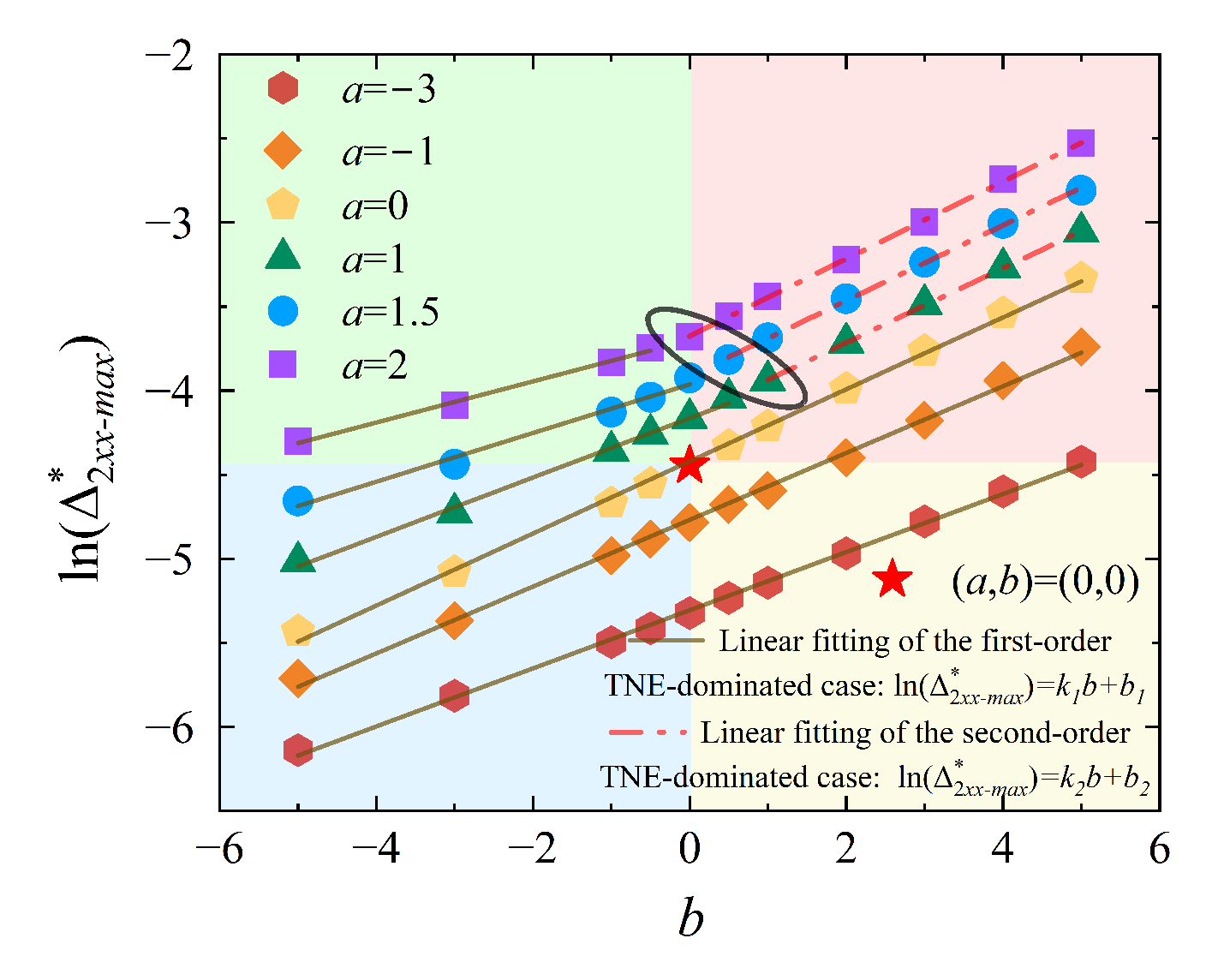}
	\caption{Phase diagram of the extremum of total viscous stress as a function of $b$ for different values of $a$. Brown solid lines denote linear fits in the first-order-TNE-dominated regime, whereas red dashed lines denote linear fits in the second-order-TNE-dominated regime.}
	\label{Fig09}
\end{figure}

%%%%%%%%%%%%%%%%%%%%%%%%%%%%%%%%%%%%%%%%%%%%%%%%%%%%%%%%%%%%%%%%%%%%
\begin{center}
\captionof{table}{Linear-fit coefficients for the phase diagram of the extremum of total viscous stress.}
\label{tab05}

\renewcommand{\arraystretch}{1.2}
\setlength{\tabcolsep}{7pt}

\begin{tabular}{>{\centering\arraybackslash}p{2.2cm}
                >{\centering\arraybackslash}p{2.0cm}
                >{\centering\arraybackslash}p{2.2cm}
                >{\centering\arraybackslash}p{2.0cm}
                >{\centering\arraybackslash}p{2.2cm}}
\hline
      & $k_1$ & $b_1$ & $k_2$ & $b_2$ \\
\hline
$a=-3$   & $0.173$ & $-5.295$ & /    & /    \\
$a=-1$   & $0.207$ & $-4.743$ & /    & /    \\
$a=0$    & $0.214$ & $-4.421$ & /    & /    \\
$a=1$    & $0.176$ & $-4.164$ & $0.221$ & $-4.157$ \\
$a=1.5$  & $0.145$ & $-3.962$ & $0.227$ & $-3.912$ \\
$a=2$    & $0.122$ & $-3.701$ & $0.230$ & $-3.676$ \\
\hline
\end{tabular}
\end{center}
%%%%%%%%%%%%%%%%%%%%%%%%%%%%%%%%%%%%%%%%%%%%%%%%%%%%%%%%%%%%%%%%%%%%

Figure~\ref{Fig09} presents the phase diagram of the extremum of total viscous stress, and table~\ref{tab05} lists the corresponding linear-fit coefficients. For each fixed $a$, the logarithm of the extremum varies approximately linearly with $b$ within each regime, implying an exponential dependence of the nonequilibrium intensity on $b$, consistent with the analytical expression for viscous stress. The brown solid lines denote linear fits in the first-order-TNE-dominated regime, with coefficients $k_1$ and $b_1$, whereas the red dashed lines denote linear fits in the second-order-TNE-dominated regime, with coefficients $k_2$ and $b_2$. The main features are discussed below.

\paragraph{(i) Combined and competing effects of $a$ and $b$}

At the reference point $(a,b)=(0,0)$, the relaxation time is constant, $\tau=\tau_0$. Relative to this point, the phase diagram can be divided into four quadrants.

In the third quadrant $(a<0,b<0)$, both exponents reduce the relaxation time relative to the reference state in the region controlling the extremum. As a result, the nonequilibrium extremum remains below the reference value, and first-order nonequilibrium effects dominate.

In the second and fourth quadrants, $(a>0,b<0)$ and $(a<0,b>0)$, the effects of $a$ and $b$ on the relaxation time oppose each other. In the second quadrant $(a>0,b<0)$, an increase in density tends to lengthen the relaxation time, whereas an increase in temperature tends to shorten it. In the fourth quadrant $(a<0,b>0)$, the trend is reversed: an increase in density tends to shorten the relaxation time, whereas an increase in temperature tends to lengthen it. In these quadrants, the nonequilibrium extremum may lie above or below the reference value, depending on which contribution is dominant.

In the first quadrant $(a>0,b>0)$, increasing both $a$ and $b$ lengthens the relaxation time and progressively enhances second-order nonequilibrium effects. The system then enters a strong relative nonequilibrium state, with $R_{TNE}$ reaching $0.35$. This indicates that the coupling between the state-dependent relaxation time and the thermodynamic gradients has entered a nonlinear regime, so that the data no longer follow a single linear relation.

\paragraph{(ii) Two-stage variation with the relaxation parameters}

Figure~\ref{Fig09} also shows that the extremum of total viscous stress depends on $b$ in a clear two-stage manner. When $a=-3,-1,0$, $\ln \left(\Delta_{2xx-\max}^{*(1)+(2)}\right)$ increases approximately linearly with $b$, indicating that first-order TNE still dominates and that the nonequilibrium response remains relatively mild. When $a=1,1.5,2$, however, a turning point appears as $b$ increases. This turning point indicates a shift in the dominant mechanism from first-order to second-order TNE, together with a transition from a near-linear response regime to a strongly nonlinear one. At the same time, the turning point moves to lower values of $b$ as $a$ increases, indicating that second-order nonequilibrium is activated more readily. Specifically, the turning point is located at approximately $b \approx 2$ for $a=1$, moves to $b \approx 1$ for $a=1.5$, and shifts further to $b \approx 0$ for $a=2$. Thus, increasing $a$ reduces the value of $b$ at which the system enters the second-order-dominated regime.

\subsubsection{Phase diagrams of the first- and second-order viscous stresses}\label{IV-1-2}

%%%%%%%%%%%%%%%%%%%%%%%%%%%%%%%
\begin{figure}\small
	\centering
	\includegraphics[width=0.98\textwidth]{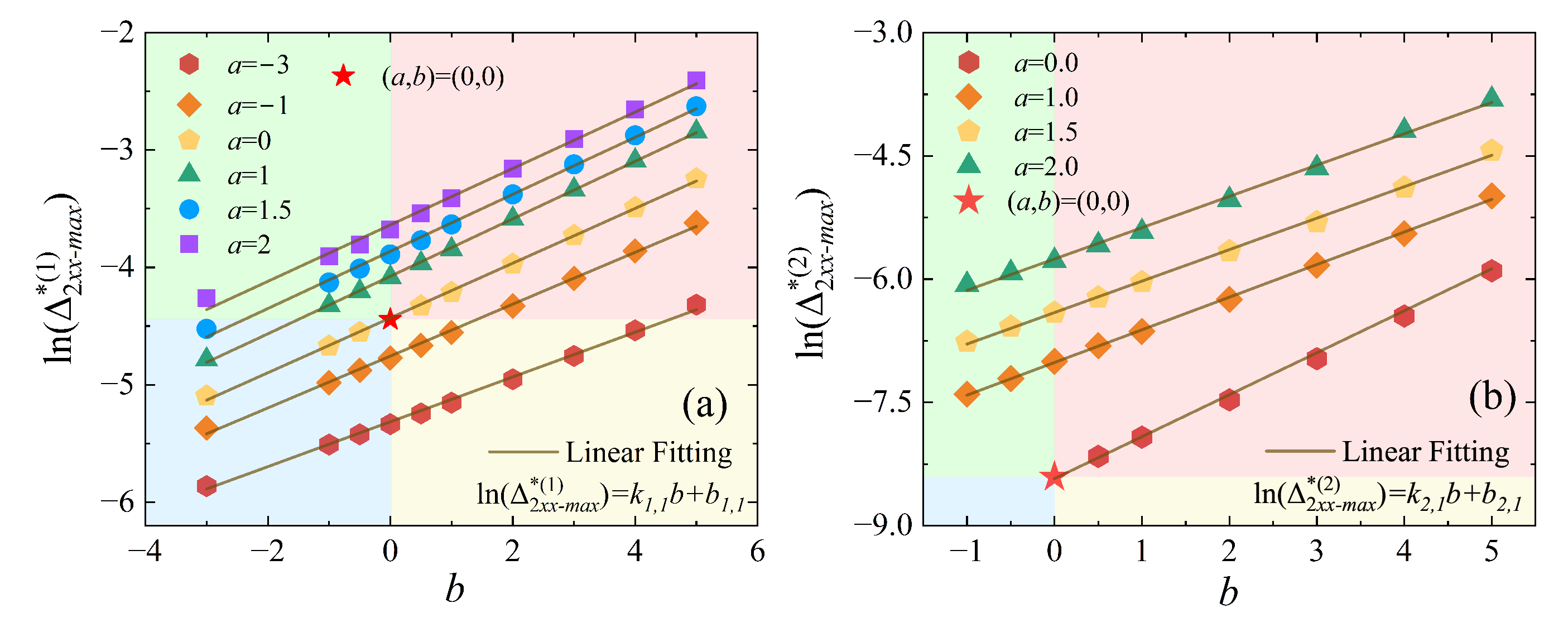}
	\caption{Phase diagrams of the logarithms of the extrema of first- and second-order viscous stresses as functions of $b$ for different values of $a$: (a) first-order viscous stress; (b) second-order viscous stress. Lines denote linear fits.}
	\label{Fig10}
\end{figure}

%%%%%%%%%%%%%%%%%%%%%%%%%%%%%%%%%%%%%%%%%%%%%%%%%%%%%%%%%%%%%%%%%%%%
\begin{center}
\captionof{table}{Linear-fit coefficients for the phase diagram of the extremum of first-order viscous stress.}
\label{tab06}

\renewcommand{\arraystretch}{1.2}
\setlength{\tabcolsep}{7pt}

\begin{tabular}{>{\centering\arraybackslash}p{3.8cm}
                >{\centering\arraybackslash}p{2.2cm}
                >{\centering\arraybackslash}p{5.6cm}}
\hline
      & $k_{1,1}$ & $b_{1,1}$ \\
\hline
$a=-3$   & $0.191$ & $-5.135$ \\
$a=-1$   & $0.221$ & $-4.755$ \\
$a=0$    & $0.233$ & $-4.431$ \\
$a=1$    & $0.243$ & $-4.075$ \\
$a=1.5$  & $0.243$ & $-3.863$ \\
$a=2$    & $0.246$ & $-3.638$ \\
\hline
\end{tabular}
\end{center}
%%%%%%%%%%%%%%%%%%%%%%%%%%%%%%%%%%%%%%%%%%%%%%%%%%%%%%%%%%%%%%%%%%%%

%%%%%%%%%%%%%%%%%%%%%%%%%%%%%%%%%%%%%%%%%%%%%%%%%%%%%%%%%%%%%%%%%%%%
\begin{center}
\captionof{table}{Linear-fit coefficients for the phase diagram of the extremum of second-order viscous stress.}
\label{tab07}

\renewcommand{\arraystretch}{1.2}
\setlength{\tabcolsep}{7pt}

\begin{tabular}{>{\centering\arraybackslash}p{3.8cm}
                >{\centering\arraybackslash}p{2.2cm}
                >{\centering\arraybackslash}p{5.6cm}}
\hline
      & $k_{2,1}$ & $b_{2,1}$ \\
\hline
$a=0$    & $0.511$ & $-8.428$ \\
$a=1$    & $0.397$ & $-7.015$ \\
$a=1.5$  & $0.383$ & $-6.407$ \\
$a=2$    & $0.375$ & $-5.754$ \\
\hline
\end{tabular}
\end{center}
%%%%%%%%%%%%%%%%%%%%%%%%%%%%%%%%%%%%%%%%%%%%%%%%%%%%%%%%%%%%%%%%%%%%

Figure~\ref{Fig10} shows the logarithms of the extrema of the first- and second-order viscous stresses as functions of $b$ for different values of $a$, together with the corresponding linear fits. Tables~\ref{tab06} and \ref{tab07} list the corresponding fit coefficients for figure~\ref{Fig10}(a,b). Several features are evident.

\paragraph{(i)} Both $\ln \left(\Delta_{2xx-\max}^{*(1)}\right)$ and $\ln \left(\Delta_{2xx-\max}^{*(2)}\right)$ increase approximately linearly with $b$, indicating that the extrema of the first- and second-order viscous stresses grow exponentially with $b$.

\paragraph{(ii)} Neither the first-order nor the second-order component exhibits clear staged behaviour. The two-stage behaviour of the total quantity must therefore arise from the coupling between contributions of different orders.

\paragraph{(iii)} The coefficient $k_{1,1}$ increases approximately linearly with $a$, for example as $k_{1,1}\approx 0.01a+0.23$. This trend indicates that first-order viscous stress becomes more sensitive to the state-dependent relaxation time as $a$ increases.

\paragraph{(iv)} By contrast, $k_{2,1}$ decreases gradually as $a$ increases. When $a=0$, the second-order viscous stress is extremely small in the low-$b$ region. Its background level is therefore low, so its relative increase with increasing $b$ appears steeper. As $a$ increases, second-order viscous stress is activated earlier and its background level rises significantly. A smaller slope does not imply weaker second-order nonequilibrium. Instead, it reflects a higher background level of second-order viscous stress. Overall, increasing $a$ and $b$ strengthens second-order nonequilibrium and progressively shifts the system from first-order dominance to second-order dominance.

Comparison of the first- and second-order phase diagrams shows that the fitted slope for second-order viscous stress is consistently larger than that for first-order viscous stress. This indicates that second-order nonequilibrium is more sensitive to $b$ and that the combined effect of $a$ and $b$ becomes more pronounced at larger $a$. Comparing the total quantity in figure~\ref{Fig09} with its components in figure~\ref{Fig10}(a,b) shows that neither component exhibits staged behaviour, whereas the total quantity does. This demonstrates that the staged behaviour of the total quantity originates from coupling between contributions of different orders.

\subsubsection{Phase diagrams of slope variation}\label{IV-1-3}

%%%%%%%%%%%%%%%%%%%%%%%%%%%%%%%
\begin{figure}\small
	\centering
	\includegraphics[width=0.65\textwidth]{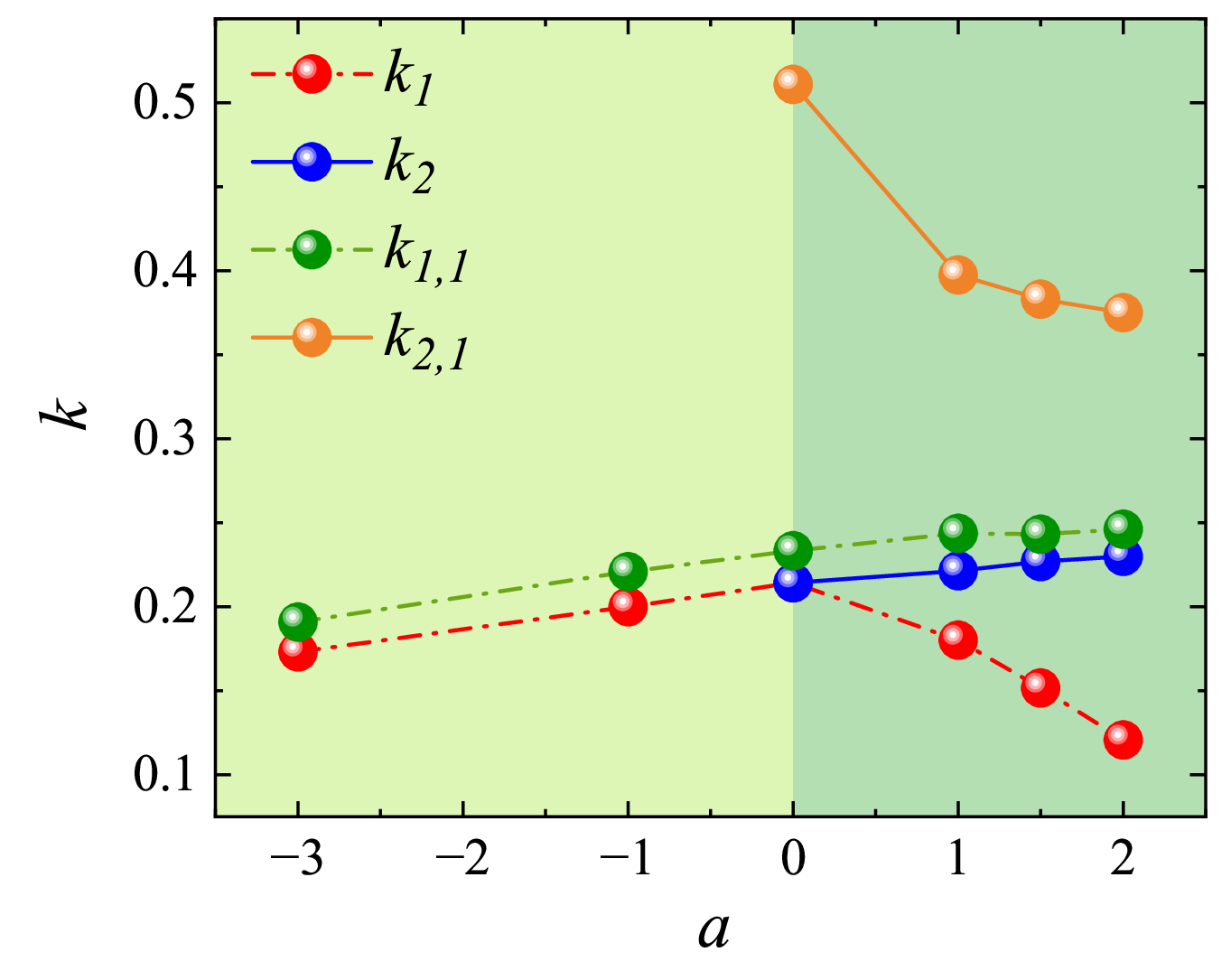}
	\caption{Variation with $a$ of the linear-fit slopes for the extrema of viscous stresses of different orders, namely $k_1$, $k_2$, $k_{1,1}$, and $k_{2,1}$.}
	\label{Fig11}
\end{figure}

Figure~\ref{Fig11} summarizes the slope information from tables~\ref{tab05}--\ref{tab07} and shows how the nonequilibrium growth rates of viscous stresses of different orders vary with $a$. Here, $k_1$ and $k_2$ denote the slopes of the total-viscous-stress phase diagram in figure~\ref{Fig09} for the first- and second-order-dominated regimes, respectively, whereas $k_{1,1}$ and $k_{2,1}$ denote the slopes of the first- and second-order viscous-stress phase diagrams in figure~\ref{Fig10}. The main observations are as follows.

\paragraph{(i) $k_{1,1}$ increases with $a$ and then gradually saturates.}
The first-order viscous stress is associated mainly with the velocity gradient. Increasing $a$ lengthens the relaxation time, enhances the nonequilibrium extremum, and therefore increases $k_{1,1}$. As the system gradually shifts from first-order to second-order dominance, however, the first-order contribution saturates and the growth of $k_{1,1}$ slows.

\paragraph{(ii) $k_{2,1}$ decreases overall as $a$ increases.}
When $a<0$, the system is dominated by first-order nonequilibrium and the second-order contribution is nearly zero. Once $a$ becomes positive, the second-order contribution is activated. As $a$ continues to increase, second-order TNE becomes stronger, but its relative growth with increasing $b$ becomes less steep, which is reflected in the gradual decrease of $k_{2,1}$.

\paragraph{(iii) The slope $k_1$ of the first-order-dominated branch of the total viscous-stress phase diagram first increases and then decreases with $a$.}
When $a<0$, first-order nonequilibrium dominates, and $k_1$ increases together with $k_{1,1}$. When $a>0$, however, $k_1$ starts to decrease. Two factors account for this behaviour.

First, the growing second-order contribution partially offsets the increase of the total extremum near the end of the first regime, thereby reducing $k_1$. Beyond the turning point, the same contribution becomes dominant and drives the second branch, so that $k_2$ remains larger than $k_1$.

Second, the relaxation time $\tau$ has a dual effect. When $a<0$, $\tau$ is small and mainly increases the nonequilibrium intensity. When $a>0$, $\tau$ becomes larger and broadens the interface more strongly, so the growth of the extremum slows and $k_1$ decreases.

\paragraph{(iv) The slope $k_2$ of the second-order-dominated branch increases slowly with $a$.}
The second-order-dominated branch is governed mainly by higher-order nonlinear transport contributions. Although $k_{2,1}$ decreases, coupling between the residual first-order contribution and the second-order term still produces a net enhancement, so that the total slope $k_2$ increases slowly. This reflects the coupling between macroscopic and mesoscopic nonequilibrium processes.

\paragraph{(v) The ratio $k_2 \approx 2k_1$ is approached only at large $a$.}
In the ideal limit of fully dominant second-order nonequilibrium, the second-order term carries exponents $2a$ and $2b$, so one expects $k_2 \approx 2k_1$. Figure~\ref{Fig11} shows that this ratio approaches $2$ only gradually at large $a$, indicating that the second-order contribution remains nonlinear and therefore deviates from the ideal limit.

\subsection{Heat flux}\label{IV-2}

We next analyse the phase-diagram characteristics of the extrema of heat-flux nonequilibrium under the same initial conditions as in section~\ref{III-2-2}.

\subsubsection{Phase diagram of the extrema of total heat flux nonequilibrium}\label{IV-2-1}

%%%%%%%%%%%%%%%%%%%%%%%%%%%%%%%
\begin{figure}\small
	\centering
	\includegraphics[width=0.80\textwidth]{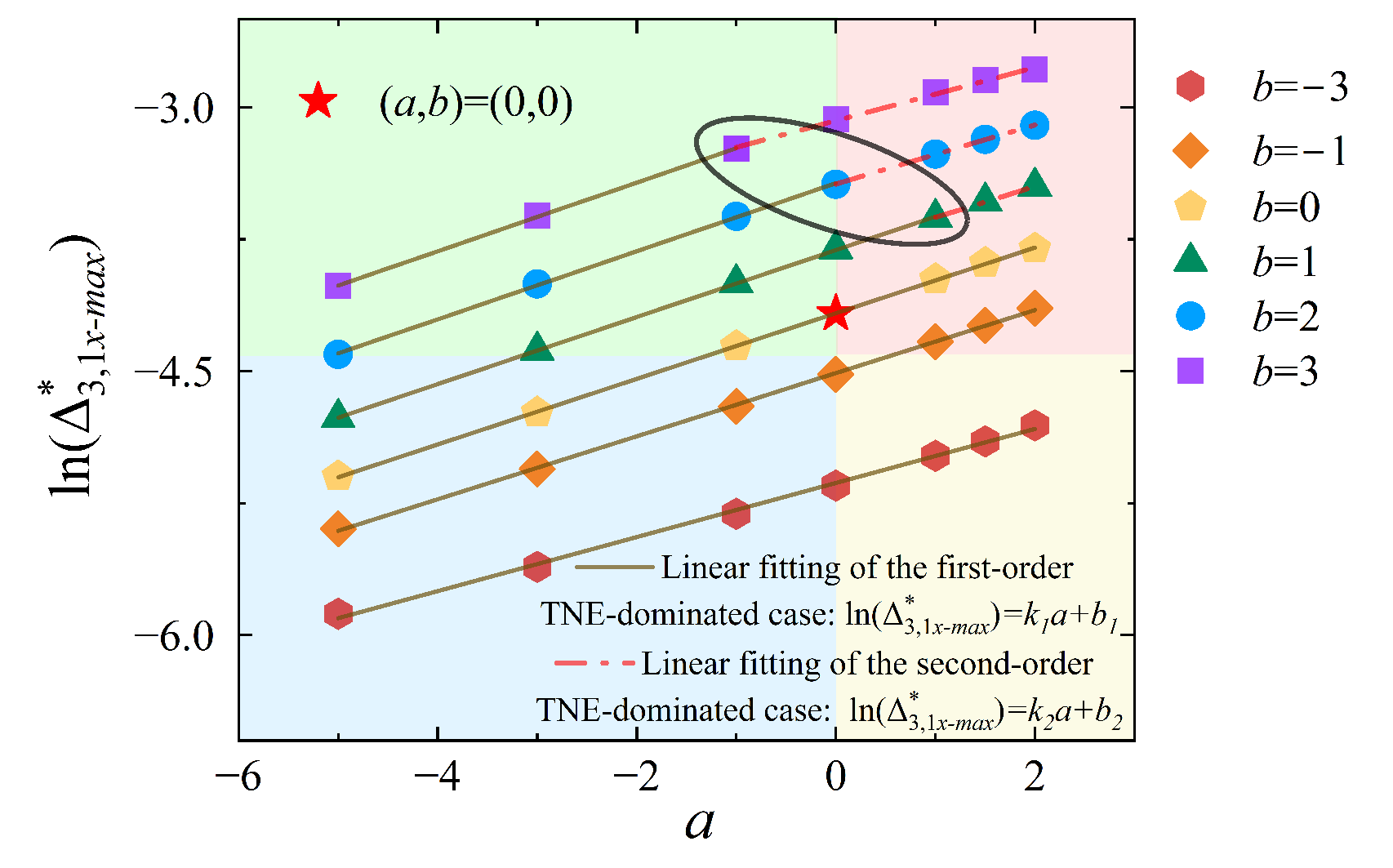}
	\caption{Phase diagram of the extremum of total heat flux as a function of $a$ for different values of $b$. Brown solid lines denote linear fits in the first-order-TNE-dominated regime, whereas red dashed lines denote linear fits in the second-order-TNE-dominated regime.}
	\label{Fig12}
\end{figure}

%%%%%%%%%%%%%%%%%%%%%%%%%%%%%%%%%%%%%%%%%%%%%%%%%%%%%%%%%%%%%%%%%%%%
\begin{center}
\captionof{table}{Linear-fit coefficients for the phase diagram of the extremum of total heat flux.}
\label{tab08}

\renewcommand{\arraystretch}{1.2}
\setlength{\tabcolsep}{7pt}

\begin{tabular}{>{\centering\arraybackslash}p{1.9cm}
                >{\centering\arraybackslash}p{2.1cm}
                >{\centering\arraybackslash}p{2.1cm}
                >{\centering\arraybackslash}p{2.1cm}
                >{\centering\arraybackslash}p{2.1cm}}
\hline
      & $k_1$ & $b_1$ & $k_2$ & $b_2$ \\
\hline
$b=-3$   & $0.154$ & $-5.136$ & /    & /    \\
$b=-1$   & $0.180$ & $-4.511$ & /    & /    \\
$b=0$    & $0.187$ & $-4.170$ & /    & /    \\
$b=1$    & $0.190$ & $-3.811$ & $0.177$ & $-3.803$ \\
$b=2$    & $0.194$ & $-3.431$ & $0.168$ & $-3.436$ \\
$b=3$    & $0.195$ & $-3.038$ & $0.151$ & $-3.076$ \\
\hline
\end{tabular}
\end{center}
%%%%%%%%%%%%%%%%%%%%%%%%%%%%%%%%%%%%%%%%%%%%%%%%%%%%%%%%%%%%%%%%%%%%

Figure~\ref{Fig12} and table~\ref{tab08} summarize the phase diagram of the extremum of total heat flux and its linear-fit coefficients. The main observations are as follows.

\paragraph{(i) Combined and competing effects of $a$ and $b$}

Taking $(a,b)=(0,0)$ as the reference point, figure~\ref{Fig12} shows that the nonequilibrium intensity exhibits distinct behaviour in the four quadrants. In the third quadrant $(a<0,b<0)$, first-order nonequilibrium effects remain dominant, and the extremum stays below the reference value. In the second and fourth quadrants, $(a>0,b<0)$ and $(a<0,b>0)$, the effects of $a$ and $b$ compete, and the nonequilibrium extremum may lie above or below the reference value. In the first quadrant $(a>0,b>0)$, the combined action of $a$ and $b$ markedly increases the extremum, and second-order heat-flux effects gradually become dominant, with $R_{TNE}$ reaching $0.58$.

\paragraph{(ii) Two-stage variation with the relaxation parameters}

As in the viscous-stress case, the extremum of total heat flux also shows a clear two-stage variation with the parameters, indicating that the system can shift from a regime dominated by first-order heat flux to one dominated by second-order heat flux. Unlike the viscous-stress case in the present configuration, however, the second-order heat flux does not continue to amplify the total nonequilibrium. Instead, once the second-order contribution becomes dominant within the parameter range examined here, the growth rate of the total-heat-flux extremum decreases. Moreover, $b$ strongly influences the location of the turning point associated with this transition. As $b$ increases, the turning point shifts towards smaller values of $a$, indicating that second-order heat-flux nonequilibrium is activated more readily: for $b=1$, the turning point is located at approximately $a \approx 1$; for $b=2$, it moves to $a \approx 0$; and for $b=3$, it shifts further to $a \approx -1$. Thus, increasing $b$ reduces the value of $a$ at which the system enters the second-order-dominated regime. Comparison of figures~\ref{Fig09} and \ref{Fig12} further shows that, although $a$ and $b$ enter the analytical expressions symmetrically, their effective influences are not the same in a given flow state. Under the present conditions, viscous stress is more sensitive to $a$, whereas heat flux is more sensitive to $b$.

\subsubsection{Phase diagrams of the first- and second-order heat flux nonequilibrium}\label{IV-2-2}
%%%%%%%%%%%%%%%%%%%%%%%%%%%%%%%
\begin{figure}\small
	\centering
	\includegraphics[width=0.98\textwidth]{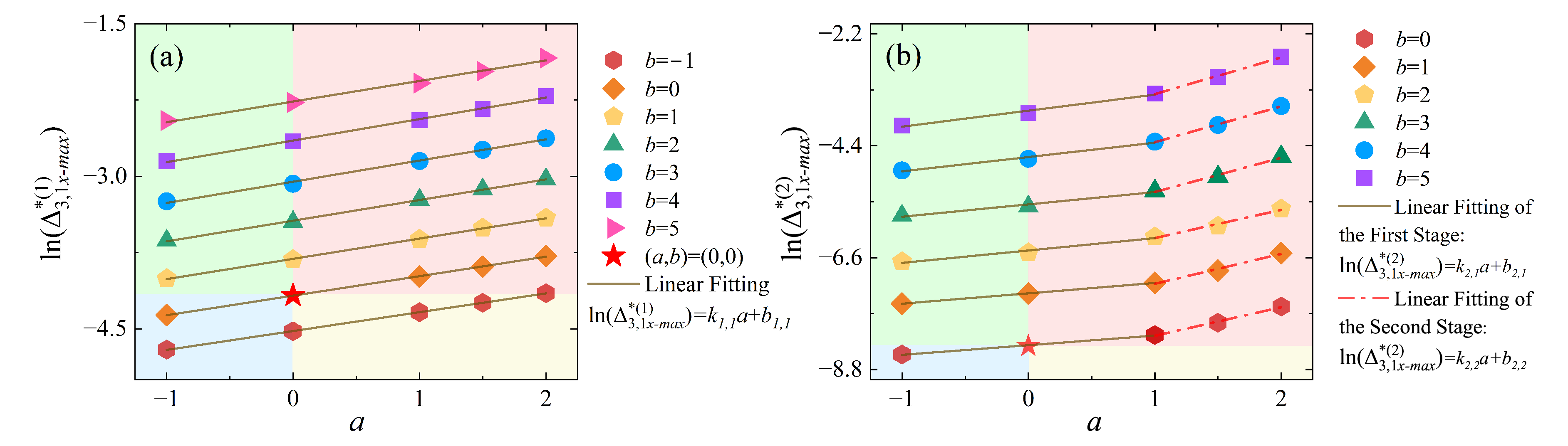}
	\caption{Phase diagrams of the extrema of first- and second-order heat fluxes as functions of $a$ for different values of $b$: (a) first-order heat flux; (b) second-order heat flux. Lines denote linear fits.}
	\label{Fig13}
\end{figure}

%%%%%%%%%%%%%%%%%%%%%%%%%%%%%%%%%%%%%%%%%%%%%%%%%%%%%%%%%%%%%%%%%%%%
\begin{center}
\captionof{table}{Linear-fit coefficients for the phase diagram of the extremum of first-order heat flux.}
\label{tab09}

\renewcommand{\arraystretch}{1.2}
\setlength{\tabcolsep}{7pt}

\begin{tabular}{>{\centering\arraybackslash}p{3.8cm}
                >{\centering\arraybackslash}p{2.2cm}
                >{\centering\arraybackslash}p{5.6cm}}
\hline
      & $k_{1,1}$ & $b_{1,1}$ \\
\hline
$b=-1$  & $0.185$ & $-4.521$ \\
$b=0$   & $0.192$ & $-4.174$ \\
$b=1$   & $0.199$ & $-3.811$ \\
$b=2$   & $0.203$ & $-3.436$ \\
$b=3$   & $0.208$ & $-3.053$ \\
$b=4$   & $0.209$ & $-2.648$ \\
$b=5$   & $0.211$ & $-2.265$ \\
\hline
\end{tabular}
\end{center}
%%%%%%%%%%%%%%%%%%%%%%%%%%%%%%%%%%%%%%%%%%%%%%%%%%%%%%%%%%%%%%%%%%%%

%%%%%%%%%%%%%%%%%%%%%%%%%%%%%%%%%%%%%%%%%%%%%%%%%%%%%%%%%%%%%%%%%%%%
\begin{center}
\captionof{table}{Linear-fit coefficients for the two-stage phase diagram of the extremum of second-order heat flux.}
\label{tab10}

\renewcommand{\arraystretch}{1.2}
\setlength{\tabcolsep}{7pt}

\begin{tabular}{>{\centering\arraybackslash}p{1.9cm}
                >{\centering\arraybackslash}p{2.1cm}
                >{\centering\arraybackslash}p{2.1cm}
                >{\centering\arraybackslash}p{2.1cm}
                >{\centering\arraybackslash}p{2.1cm}}
\hline
      & $k_{2,1}$ & $b_{2,1}$ & $k_{2,2}$ & $b_{2,2}$ \\
\hline
$b=0$  & $0.190$ & $-8.322$ & $0.530$ & $-8.696$ \\
$b=1$  & $0.204$ & $-7.301$ & $0.547$ & $-7.700$ \\
$b=2$  & $0.224$ & $-6.458$ & $0.553$ & $-6.767$ \\
$b=3$  & $0.242$ & $-5.552$ & $0.672$ & $-5.980$ \\
$b=4$  & $0.282$ & $-4.621$ & $0.703$ & $-5.030$ \\
$b=5$  & $0.314$ & $-3.709$ & $0.721$ & $-4.105$ \\
\hline
\end{tabular}
\end{center}
%%%%%%%%%%%%%%%%%%%%%%%%%%%%%%%%%%%%%%%%%%%%%%%%%%%%%%%%%%%%%%%%%%%%

\paragraph{(i) Trend of the first-order heat flux}

Figure~\ref{Fig13}(a) shows the logarithm of the extremum of first-order heat flux as a function of $a$ for different values of $b$, and table~\ref{tab09} lists the corresponding fit coefficients. The results show that $\ln \left(\Delta_{3,1x-\max}^{*(1)}\right)$ increases approximately linearly with $a$, indicating that the first-order heat flux grows exponentially with $a$. The curves depend only weakly on $b$, and the fitted slope increases only slightly, so the overall growth remains smooth and stable.

\paragraph{(ii) Two-stage variation of the second-order heat flux}

Figure~\ref{Fig13}(b) shows the logarithm of the extremum of second-order heat flux as a function of $a$ for different values of $b$, and table~\ref{tab10} lists the corresponding fit coefficients. Although $\ln \left(\Delta_{3,1x-\max}^{*(2)}\right)$ varies approximately linearly with $a$ overall, a clear two-stage behaviour appears in the later regime, where the slope increases markedly. This indicates that increasing $a$ accelerates the growth of second-order heat flux and progressively makes it dominant.

\paragraph{(iii) Coupling effects and the decrease in the growth rate of total heat flux}

Together with figures~\ref{Fig12} and \ref{Fig13}(a,b), these results show that the staged behaviour of total heat-flux nonequilibrium originates from the competitive coupling between the first- and second-order heat fluxes. In the first-order-dominated stage, the total heat flux increases approximately linearly, reflecting ordinary gradient-driven energy transport. Once the second-order contribution becomes appreciable, nonlinear coupling is strengthened. Although the second-order heat flux grows more rapidly, the growth rate of the total heat flux decreases.

\subsubsection{Phase diagrams of slope variation}\label{IV-2-3}
%%%%%%%%%%%%%%%%%%%%%%%%%%%%%%%
\begin{figure}\small
	\centering
	\includegraphics[width=0.65\textwidth]{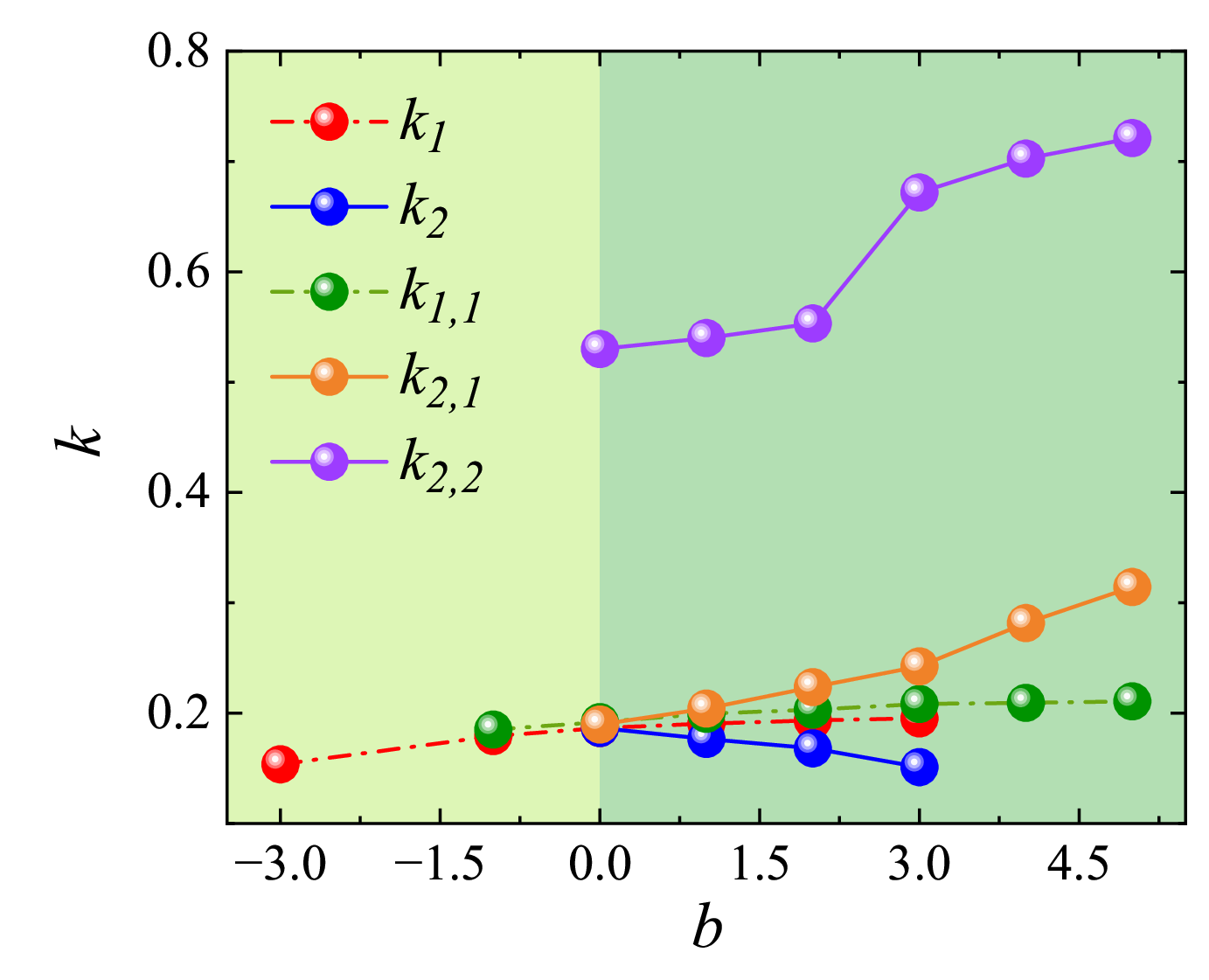}
	\caption{Variation with $b$ of the linear-fit slopes for the extrema of heat fluxes of different orders, namely $k_1$, $k_2$, $k_{1,1}$, $k_{2,1}$, and $k_{2,2}$.}
	\label{Fig14}
\end{figure}

Figure~\ref{Fig14} summarizes the slope information from tables~\ref{tab08}--\ref{tab10} and shows how the nonequilibrium growth rates of heat fluxes of different orders evolve with $b$. Here, $k_1$ and $k_2$ denote the slopes of the total-heat-flux phase diagram in figure~\ref{Fig12} for the first- and second-order-dominated regimes, respectively. The coefficient $k_{1,1}$ is the slope of the first-order heat-flux phase diagram in figure~\ref{Fig13}(a), whereas $k_{2,1}$ and $k_{2,2}$ are the slopes of the two stages in the second-order heat-flux phase diagram in figure~\ref{Fig13}(b).

\paragraph{(i) The first-order heat-flux slope $k_{1,1}$ remains nearly constant.}
The weak dependence of $k_{1,1}$ on $b$ indicates that first-order heat flux is governed mainly by gradient-driven energy transport.

\paragraph{(ii) The slope $k_1$ of the first-order-dominated branch increases slightly with $b$ and then approaches a plateau.}
As $b$ increases, the relaxation time becomes longer, and weak second-order effects are partially activated, acting cooperatively with the first-order term. However, no strong additional contribution has yet formed, and the system remains in a nearly linear regime.

\paragraph{(iii) The slope $k_2$ of the second-order-dominated branch decreases as $b$ increases.}
Unlike the viscous-stress case, $k_2$ for heat flux decreases with increasing $b$. This indicates that, for the heat-flux configuration considered here, second-order nonlinear coupling strengthens in the strongly nonequilibrium regime. Although the nonequilibrium intensity continues to increase, its growth rate decreases as the second-order heat-flux contribution becomes stronger.

\paragraph{(iv) The slopes $k_{2,1}$ and $k_{2,2}$ of second-order heat flux both increase with $b$.}
As $b$ increases, second-order nonequilibrium becomes more pronounced, and both $k_{2,1}$ and $k_{2,2}$ increase accordingly. When $a$ is small, the second-order contribution has not yet fully developed. When $a$ is large, nonlinear coupling is fully activated, leading to $k_{2,2}>k_{2,1}$. In this regime, the combined effect of $a$ and $b$ becomes more evident in the evolution of heat-flux nonequilibrium.

\subsection{Summary}\label{IV-3}

The phase diagrams of viscous stress and heat flux show how the relaxation parameters $a$ and $b$ affect nonequilibrium transport in a systematic and channel-dependent manner. Through the state-dependent power-law relaxation time, different parameter combinations lead to different nonequilibrium intensities and can progressively shift the system from a first-order linear nonequilibrium regime to a second-order nonlinear one within the parameter range considered here. The two transport channels, however, respond differently. Under the present conditions, the second-order viscous-stress contribution increases the growth rate of the extremum of the total viscous stress, whereas the second-order heat-flux contribution reduces the growth rate of the extremum of the total heat flux. Higher-order nonequilibrium contributions therefore affect the overall nonequilibrium response in different ways in different transport channels.

This difference in behaviour does not imply that higher-order viscous stress simply enhances nonequilibrium whereas higher-order heat flux suppresses it. Instead, it originates from the different ways in which the higher-order momentum flux and higher-order energy flux act back on their respective driving gradients. In the viscous-stress case considered here, the higher-order momentum flux is significantly amplified by the $\tau^2$-type state-dependent terms and is activated mainly in the interfacial region where the first-order viscous-stress peak is located. It is therefore co-located with the original nonequilibrium momentum flux. Within the present parameter range, this amplification effect outweighs the weakening caused by velocity-gradient smoothing, leading to an increased growth rate of the extremum of the total viscous stress. By contrast, in the heat-flux case, although the higher-order energy flux also increases with the relaxation parameters, its enhancement more directly promotes thermal redistribution and smoothing of the local temperature gradient, thereby weakening the $\bm{\nabla} T$ that drives the continued rapid growth of the heat flux. As a result, the extremum of the total heat flux still increases with the relaxation parameters, but its growth rate progressively decreases.

\section{Conclusions and outlook}\label{V}

Within the framework of the discrete Boltzmann method (DBM), we have developed a discrete Boltzmann model with a density- and temperature-dependent power-law relaxation time (DTRT-DBM), in which the constant relaxation time of the conventional BGK model is generalized to a state-dependent form. This formulation allows the relaxation rate to vary with density and temperature while retaining the structural simplicity and physical interpretability of DBM. It thereby establishes an explicit coupling between macroscopic states, mesoscopic relaxation, and nonequilibrium transport processes, and provides a framework for modelling nonequilibrium flows with spatially varying local states.

One-dimensional compressible-flow benchmarks show that the model accurately recovers macroscopic conservation laws, remains numerically stable, and captures nonequilibrium quantities effectively. On this basis, we construct phase diagrams of nonequilibrium extrema and growth slopes for two representative quantities, viscous stress and heat flux, and systematically analyse how the model parameters $a$ and $b$ affect the intensity and structural variation of nonequilibrium. The results show that the nonequilibrium peaks increase exponentially with the model parameters and exhibit clear regime-dependent behaviour. Under suitable conditions, the system shifts from first-order linear dominance to second-order nonlinear dominance, highlighting the cross-scale nonequilibrium characteristics captured by the power-law-dependent relaxation-time formulation.

The nonequilibrium phase diagrams further show that the relaxation parameters $a$ and $b$ jointly affect both viscous stress and heat flux, and that both quantities depend exponentially on these parameters. When both parameters are sufficiently large, the extrema display clear stage-dependent variation, indicating a shift in the governing nonequilibrium transport mechanism from low-order to high-order dominance. More specifically, under the flow configurations and parameter range considered here, the emergence of second-order viscous stress increases the growth rate of the total viscous-stress extremum, whereas the emergence of second-order heat flux reduces the growth rate of the total heat-flux extremum. These results show that, within the present DTRT-DBM framework and for the conditions considered here, higher-order nonequilibrium contributions affect different transport channels in different ways.

At present, the model considers only the dependence of $\tau$ on $\rho$ and $T$, and the validation and mechanistic analysis remain focused mainly on one-dimensional cases. In more complex settings, the relaxation process may also be influenced by pressure, velocity shear, strain rate, or coupling to external forces \citep{li2020ICHMT, li2020IJNFM, wu2021JCP}. Incorporating these effects should further enhance the generality and predictive capability of the model.

Future work may proceed in two directions. One is to develop more general dynamic relaxation-time forms with multi-parameter coupling, and to combine them with more complex collision models or source terms associated with external fields, so as to achieve a more detailed description of multiple dissipation mechanisms. The other is to extend the DTRT-DBM to two- and three-dimensional strong-gradient flows and coupled multiphysics configurations, where its ability to recover realistic transport processes and capture nonequilibrium effects across scales can be further assessed in physically relevant settings.
\linespread{1.0}

\section*{Acknowledgements}
This work was supported by the National Natural Science Foundation of China (Grant Nos. U2242214 and 11875001), the Hebei Outstanding Youth Science Foundation (Grant No. A2023409003), the Natural Science Foundation of Fujian Province (Grant No. 2026J001415), and the Fujian Provincial Special Funds for Education and Research in Provincial Units (Grant No. K3-949).

\appendix
\section{Kinetic Moment Relations}\label{A}

\begin{equation}\label{A1}
\mathbf{M}_0=\sum_i f_i^{e q}=\rho
\end{equation}

\begin{equation}\label{A2}
\mathbf{M}_1=\sum_i f_i^{e q} \mathbf{v}_i=\rho \mathbf{u},
\end{equation}

\begin{equation}\label{A3}
\mathbf{M}_{2,0}=\sum_i \frac{1}{2} f_i^{e q}\left(v_i^2+\eta_i^2\right)=\frac{1}{2} \rho\left[(n+2) R T+u^2\right]
\end{equation}

\begin{equation}\label{A4}
\mathbf{M}_2=\sum_i f^{e q} \mathbf{v}_i \mathbf{v}_i=\rho(R T \mathbf{I}+\mathbf{u u}),
\end{equation}

\begin{equation}\label{A5}
\mathbf{M}_{3,1}=\sum_i \frac{1}{2} f_i^{e q}\left(v_i^2+\eta_i^2\right) \mathbf{v}_i=\frac{1}{2} \rho \mathbf{u}\left[(n+4) R T+u^2\right],
\end{equation}

\begin{equation}\label{A6}
\mathbf{M}_3=\sum_i f_i^{e q} \mathbf{v}_i \mathbf{v}_i \mathbf{v}_i=\rho\left[\left(u_\alpha \delta_{\beta \gamma}+u_\beta \delta_{\alpha \gamma}+u_\gamma \delta_{\alpha \beta}\right) \mathbf{e}_\alpha \mathbf{e}_\beta \mathbf{e}_\gamma R T+\mathbf{u u u}\right],
\end{equation}

\begin{equation}\label{A7}
\mathbf{M}_{4,2}=\sum_i \frac{1}{2} f_i^{e q}\left(v_i^2+\eta_i^2\right) \mathbf{v}_i \mathbf{v}_i=\rho\left[\left(\frac{n+4}{2} R T+\frac{u^2}{2}\right) R T \mathbf{I}+\left(\frac{n+6}{2} R T+\frac{u^2}{2}\right) \mathbf{u u}\right],
\end{equation}

\begin{equation}\label{A8}
\begin{aligned}
\mathbf{M}_4 & =\sum_i f_i^{e q} \mathbf{v}_i \mathbf{v}_i \mathbf{v}_i \mathbf{v}_i=\rho\left[R^2 T^2\left(\delta_{\alpha \beta} \delta_{\gamma \lambda}+\delta_{\alpha \gamma} \delta_{\beta \lambda}+\delta_{\alpha \lambda} \delta_{\beta \gamma}\right) \mathbf{e}_\alpha \mathbf{e}_\beta \mathbf{e}_\gamma \mathbf{e}_\lambda\right. \\
& +R T\left(u_\alpha u_\beta \delta_{\gamma \lambda}+u_\alpha u_\gamma \delta_{\beta \lambda}+u_\alpha u_\lambda \delta_{\beta \gamma}+u_\beta u_\gamma \delta_{\alpha \lambda}+u_\beta u_\lambda \delta_{\alpha \gamma}\right. \\
& \left.\left.+u_\gamma u_\lambda \delta_{\alpha \beta}\right) \mathbf{e}_\alpha \mathbf{e}_\beta \mathbf{e}_\gamma \mathbf{e}_\lambda+\mathbf{u u u u}\right],
\end{aligned}
\end{equation}

\begin{equation}\label{A9}
\begin{aligned}
\mathbf{M}_{5,3}= & \sum_i \frac{1}{2} f_i^{e q}
\left(v_i^2+\eta_i^2\right)
\mathbf{v}_i \mathbf{v}_i \mathbf{v}_i
=\rho\left[
\left(\frac{n+8}{2} R T+\frac{u^2}{2}\right)
\mathbf{u} \mathbf{u} \mathbf{u}
\right. \\
&\quad \left.
+\left(\frac{n+6}{2} R T+\frac{u^2}{2}\right)
\left(u_\alpha \delta_{\beta \gamma}
+u_\beta \delta_{\alpha \gamma}
+u_\gamma \delta_{\alpha \beta}\right)
\mathbf{e}_\alpha \mathbf{e}_\beta \mathbf{e}_\gamma R T
\right].
\end{aligned}
\end{equation}

\section{Second-order Constitutive Relations}\label{B}

For the DTRT model, the second-order components of the non-organized momentum flux $\bm{\Delta}_2^*$ and the non-organized energy flux $\bm{\Delta}_{3,1}^*$ are given by

\begin{equation}\label{B1}
\Delta_{2 x x}^{*(2)}
=-\frac{2 R \tau_0^2}{(n+2)^2 \rho}
\left(\frac{\rho}{\rho_0}\right)^{2 a}
\left(\frac{T}{T_0}\right)^{2 b}
\left(F_1+F_2+F_3\right),
\end{equation}
\begin{equation}\label{B2}
\Delta_{2 x y}^{*(2)}
=\tau_0^2
\left(\frac{\rho}{\rho_0}\right)^{2 a}
\left(\frac{T}{T_0}\right)^{2 b}
\left(-\frac{\rho R T}{n+2} F_4+F_5\right),
\end{equation}
\begin{equation}\label{B3}
\Delta_{2 y y}^{*(2)}
=\frac{2 R \tau_0^2}{(n+2)^2 \rho}
\left(\frac{\rho}{\rho_0}\right)^{2 a}
\left(\frac{T}{T_0}\right)^{2 b}
\left(F_6+F_7+F_8\right),
\end{equation}
\begin{equation}\label{B4}
\Delta_{3,1 x}^{*(2)}
=\frac{\tau_0^2 R^2 T}{n+2}
\left(\frac{\rho}{\rho_0}\right)^{2 a}
\left(\frac{T}{T_0}\right)^{2 b}
\left(F_9+F_{10}+F_{11}+F_{12}+F_{13}\right),
\end{equation}
\begin{equation}\label{B5}
\Delta_{3,1 y}^{*(2)}
=\frac{4 \tau_0^2 R^2 T}{n+2}
\left(\frac{\rho}{\rho_0}\right)^{2 a}
\left(\frac{T}{T_0}\right)^{2 b}
\left(F_{14}+F_{15}+F_{16}+F_{17}+F_{18}\right),
\end{equation}

The expressions for $F_{1}-F_{18}$ are listed in table~\ref{tab00}.

%%%%%%%%%%%%%%%%%%%%%%%%%%%%%%%%%%%%%%%%%%%%%%%%%%%%%%%%%%%%%%%%%%%%
% ---------- longtable ���У����������Զ��Գƣ�----------
\setlength{\LTleft}{\fill}
\setlength{\LTright}{\fill}

% �иߡ��м��ࣨ�����������ã���΢����
\renewcommand{\arraystretch}{1.25}
\setlength{\tabcolsep}{2pt}  % ? �� 6pt ��խ���ܿ��ȸ�С�����ᳬ��ҳü��

% ? �ڶ��п��ȣ�������խ�ͼ�����С��0.82 -> 0.80��
%\newlength{\ExprColW}
%\setlength{\ExprColW}{0.84\textwidth}
\makeatletter
\@ifundefined{ExprColW}{\newlength{\ExprColW}}{}
\makeatother
\setlength{\ExprColW}{0.84\textwidth}

\begin{longtable}{@{}
>{\centering\arraybackslash}m{2.0cm}
>{\centering\arraybackslash}p{\ExprColW}
@{}}

\caption{Grouped analytical expressions for the second-order constitutive terms in the DTRT-DBM}
\label{tab00}\\

\toprule
No. & \multicolumn{1}{c}{Analytical expression} \\
\midrule
\endfirsthead

% ---------- ����һҳ�ײ����� ----------
\bottomrule
\endlastfoot

% ===================== ���濪ʼ���ģ����Ĺ�ʽ�ѱ����� =====================

$F_{1}$ &
\(
\displaystyle
\begin{aligned}[t]
& -2(n+1)\rho^2 T u_x \partial^2_{x} u_x
-n\rho^2 T u_y \partial^2_{xy} u_x
-(n+2)\rho^2 T u_x \partial^2_{y} u_x \\
&\quad
-(n+2)\rho^2 T u_y \partial^2_{x} u_y
-n\rho^2 T u_x \partial^2_{xy} u_y
-2(n+1)\rho^2 T u_y \partial^2_{y} u_y \\
&\quad
+(n+1)(n+2)\rho R T^2 \partial^2_{x} \rho
-(n+2)\rho R T^2 \partial^2_{y} \rho \\
&\quad
-(n+1)(n+2)R T^2 \bigl(\partial_x \rho\bigr)^2
+(n+2)R T^2 \bigl(\partial_y \rho\bigr)^2
\end{aligned}
\)
\\[2pt]
\cmidrule(lr){1-2}

$F_{2}$ &
\(
\displaystyle
\begin{aligned}[t]
& -\rho T
\biggl\{
(n+2)(a+1)u_y \partial_x u_y
-2(a+1)u_x \partial_y u_y \\
&\quad
+(n+2)(a+1)u_y \partial_y u_x
+(n+1)\partial_x \rho
\Bigl[(n+2)aR \partial_x T \\
&\quad
+2(a+1)u_x \partial_x u_x\Bigr]
\biggr\} \\
&\quad
+\rho T
\biggl\{
-(n+2)(a+1)u_x \partial_x u_y
+a(n+2)R \partial_y T \\
&\quad
-(a+1)
\Bigl[2(n+1)u_y \partial_y u_y
+(n+2)u_x \partial_y u_x
-2u_y \partial_x u_x\Bigr]
\biggr\}\partial_y \rho
\end{aligned}
\)
\\[2pt]
\cmidrule(lr){1-2}

$F_{3}$ &
\(
\displaystyle
\begin{aligned}[t]
& -\rho^2
\biggl\{
2(n+2)
\Bigl[\tfrac{1}{2}(b+1)\partial_y T u_x
+\tfrac{1}{2}(b+1)\partial_x T u_y
+T \partial_y u_x\Bigr]\partial_x u_y \\
&\quad
-(n+2)(b+1)R \bigl(\partial_y T\bigr)^2 \\
&\quad
+(b+1)
\Bigl[2(n+1)u_y \partial_y u_y
+(n+2)u_x \partial_y u_x
-2u_y \partial_x u_x\Bigr]\partial_y T \\
&\quad
+\Bigl[n(a+1)+2a+2b+4\Bigr]T \bigl(\partial_y u_y\bigr)^2 \\
&\quad
+\Bigl[
-2(b+1)\partial_x T u_x
-\bigl(4+a n^2+\bigl[2a+2b+4\bigr]n\bigr)T \partial_x u_x
\Bigr]\partial_y u_y \\
&\quad
+(n+2)^2T \bigl(\partial_y u_x\bigr)^2
+(n+2)(b+1)\partial_x T u_y \partial_y u_x \\
&\quad
+(n+1)
\Bigl[
(n+2)(b+1)R \bigl(\partial_x T\bigr)^2
+2(b+1)\partial_x T u_x \partial_x u_x \\
&\quad
-\bigl(\bigl[a-1\bigr]n+2a+2b\bigr)T \bigl(\partial_x u_x\bigr)^2
\Bigr]
\biggr\}
\end{aligned}
\)
\\[2pt]
\cmidrule(lr){1-2}

$F_{4}$ &
\(
\displaystyle
\begin{aligned}[t]
& \Bigl[
(na+2a+2b+4)\partial_y u_x
+(na+2a+2b-2n)\partial_x u_y
\Bigr]\partial_x u_x \\
&\quad
+(na+2a+2b-2n)\partial_y u_y \partial_y u_x
+(na+2a+2b+4)\partial_y u_y \partial_x u_y
\end{aligned}
\)
\\[2pt]
\cmidrule(lr){1-2}

$F_{5}$ &
\(
\displaystyle
\begin{aligned}[t]
& 2(b+1)\rho R^2\partial_x T \partial_y T
-2R^2T^2\partial^2_{xy} \rho
+aR^2T\bigl(\partial_x \rho \partial_y T+\partial_y \rho \partial_x T\bigr) \\
&\quad
+2R^2T^2\frac{\partial_x \rho \partial_y \rho}{\rho}
\end{aligned}
\)
\\[2pt]
\cmidrule(lr){1-2}

$F_{6}$ &
\(
\displaystyle
\begin{aligned}[t]
& 2(n+1)\rho^2T u_x \partial^2_{x} u_x
+n\rho^2T u_y \partial^2_{xy} u_x
+(n+2)\rho^2T u_x \partial^2_{y} u_x \\
&\quad
+(n+2)\rho^2T u_y \partial^2_{x} u_y
+n\rho^2T u_x \partial^2_{xy} u_y
+2(n+1)\rho^2T u_y \partial^2_{y} u_y \\
&\quad
+(n+2)\rho R T^2\partial^2_{x} \rho
-(n+2)(n+1)\rho R T^2\partial^2_{y} \rho \\
&\quad
-(n+2)R T^2\bigl(\partial_x \rho\bigr)^2
+(n+2)(n+1)R T^2\bigl(\partial_y \rho\bigr)^2
\end{aligned}
\)
\\[2pt]
\cmidrule(lr){1-2}

$F_{7}$ &
\(
\displaystyle
\begin{aligned}[t]
& -\rho T
\biggl\{
-(n+2)(a+1)u_y \partial_y u_x
+a(n+2)R \partial_x T \\
&\quad
-2(a+1)
\Bigl[(n+1)u_x \partial_x u_x
+\tfrac{1}{2}(n+2)u_y \partial_x u_y
-u_x \partial_y u_y\Bigr]
\biggr\}\partial_x \rho \\
&\quad
+\rho T
\biggl\{
(n+2)(a+1)u_x \partial_y u_x
-2(a+1)u_y \partial_x u_x \\
&\quad
+(n+2)(a+1)u_x \partial_x u_y
+(n+1)
\Bigl[a(n+2)R \partial_y T \\
&\quad
+2(a+1)u_y \partial_y u_y\Bigr]
\biggr\}\partial_y \rho
\end{aligned}
\)
\\[2pt]
\cmidrule(lr){1-2}

$F_{8}$ &
\(
\displaystyle
\begin{aligned}[t]
& -\rho^2
\biggl\{
-2(n+2)
\Bigl[\tfrac{1}{2}(b+1)\partial_x T u_y
+\tfrac{1}{2}(b+1)\partial_y T u_x
+T \partial_x u_y\Bigr]\partial_y u_x \\
&\quad
+(n+2)(b+1)R\bigl(\partial_x T\bigr)^2 \\
&\quad
-(b+1)
\Bigl[2(n+1)u_x \partial_x u_x
+(n+2)u_y \partial_x u_y
-2u_x \partial_y u_y\Bigr]\partial_x T \\
&\quad
-\Bigl[(a+1)n+2a+2b+4\Bigr]T \bigl(\partial_x u_x\bigr)^2 \\
&\quad
+\Bigl[
2(b+1)\partial_y T u_y
+\bigl(4+a n^2+\bigl[2a+2b+4\bigr]n\bigr)T \partial_y u_y
\Bigr]\partial_x u_x \\
&\quad
-(n+2)^2T \bigl(\partial_x u_y\bigr)^2
-(n+2)(b+1)\partial_y T u_x \partial_x u_y \\
&\quad
-(n+1)
\Bigl[
(n+2)(b+1)R\bigl(\partial_y T\bigr)^2
+2(b+1)\partial_y T u_y \partial_y u_y \\
&\quad
-\bigl(\bigl[a-1\bigr]n+2a+2b\bigr)T \bigl(\partial_y u_y\bigr)^2
\Bigr]
\biggr\}
\end{aligned}
\)
\\[2pt]
\cmidrule(lr){1-2}

$F_{9}$ &
\(
\displaystyle
\begin{aligned}[t]
& \rho T
\Bigl[
(n+2)\partial^2_{x} u_y
+(n-2)\partial^2_{y} u_y
-4\partial^2_{xy} u_x
\Bigr]
\end{aligned}
\)
\\[2pt]
\cmidrule(lr){1-2}

$F_{10}$ &
\(
\displaystyle
\begin{aligned}[t]
& \frac{1}{2}\rho
\biggl\{
\Bigl[a n^2+(6a+2b+4)n+8a+12b+24\Bigr]\partial_x u_x \\
&\quad
+\Bigl[(a-2)n^2+(6a-2b-16)n+8a+4b-8\Bigr]\partial_y u_y
\biggr\}\partial_y T
\end{aligned}
\)
\\[2pt]
\cmidrule(lr){1-2}

$F_{11}$ &
\(
\displaystyle
\begin{aligned}[t]
& -2aT(n+1)\partial_y \rho \partial_y u_y
\end{aligned}
\)
\\[2pt]
\cmidrule(lr){1-2}

$F_{12}$ &
\(
\displaystyle
\begin{aligned}[t]
& -\frac{1}{2}\rho(n+2)
\Bigl[
(b+2)\partial_y u_x
+(b+n+6)\partial_x u_y
\Bigr]\partial_x T
\end{aligned}
\)
\\[2pt]
\cmidrule(lr){1-2}

$F_{13}$ &
\(
\displaystyle
\begin{aligned}[t]
& -\frac{aT}{2}
\Bigl[
-2\partial_y \rho \partial_x u_x
+\partial_x \rho (n+2)\bigl(\partial_y u_x+\partial_x u_y\bigr)
\Bigr]
\end{aligned}
\)
\\[2pt]
\cmidrule(lr){1-2}

$F_{14}$ &
\(
\displaystyle
\begin{aligned}[t]
& \rho T
\Bigl[
-\frac{1}{4}(n+2)\partial^2_{x} u_y
-\frac{1}{4}(n-2)\partial^2_{y} u_y
+\partial^2_{xy} u_x
\Bigr]
\end{aligned}
\)
\\[2pt]
\cmidrule(lr){1-2}

$F_{15}$ &
\(
\displaystyle
\begin{aligned}[t]
& \frac{1}{8}\rho
\biggl\{
\Bigl[(a-2)n^2+(6a-2b-16)n+8a+4b-8\Bigr]\partial_y u_y \\
&\quad
+\Bigl[a n^2+(6a+2b+4)n+8a+12b+24\Bigr]\partial_x u_x
\biggr\}\partial_y T
\end{aligned}
\)
\\[2pt]
\cmidrule(lr){1-2}

$F_{16}$ &
\(
\displaystyle
\begin{aligned}[t]
& -\frac{1}{2}aT \partial_y \rho (n+1)\partial_y u_y
\end{aligned}
\)
\\[2pt]
\cmidrule(lr){1-2}

$F_{17}$ &
\(
\displaystyle
\begin{aligned}[t]
& -\frac{1}{4}(n+2)\rho
\Bigl[
(b+2)\partial_y u_x
+(b+n+6)\partial_x u_y
\Bigr]\partial_x T
\end{aligned}
\)
\\[2pt]
\cmidrule(lr){1-2}

$F_{18}$ &
\(
\displaystyle
\begin{aligned}[t]
& -\frac{1}{4}aT
\Bigl[
-2\partial_y \rho \partial_x u_x
+\partial_x \rho (n+2)\bigl(\partial_y u_x+\partial_x u_y\bigr)
\Bigr]
\end{aligned}
\)
\\[2pt]

\end{longtable}

% ---------- �ָ�Ĭ�ϣ���ѡ��----------
\setlength{\LTleft}{0pt}
\setlength{\LTright}{0pt}
%%%%%%%%%%%%%%%%%%%%%%%%%%%%%%%%%%%%%%%%%%%%%%%%%%%%%%%%%%%%%%%%%%%%

\section*{Declaration of interests}
The authors report no conflict of interest.

\section*{Author ORCIDs}
D. Li, https://orcid.org/0000-0003-3726-9815

Z. He, https://orcid.org/0009-0000-7232-8329

H. Lai, https://orcid.org/0000-0001-5978-5736

Y. Gan, https://orcid.org/0000-0002-0191-9022

H. Liu, https://orcid.org/0000-0002-8780-0398

P. Lin, https://orcid.org/0000-0003-2361-0066

% ��������ʦ��ORCID

%\FloatBarrier

%%%%%%%%%%%%%%%%%%%%%%%%%%%%%%%%%%%%%%%%%%%%%%%%%%%%%%%%%%%%%%%%%%
\FloatBarrier

% Note the spaces between the initials
%\newpage
\bibliographystyle{jfm}
\bibliography{References}
\end{document}